\documentstyle[prd,aps,psfig]{revtex}
\begin{document}
\def\Universita{Universit\'a}
\def\Paris{Par\'{\i}s}
\def\Perez{P\'erez}
\def\Gunther{G\"unther}
\def\Schutzhold{Sch\"utzhold}
\def\Lofstedt{L\"{o}fstedt}
\def\Garcia{Garc\'\i{}a}
\def\Ruggeberg{R\"uggeberg}
\title{Sonoluminescence as a QED vacuum effect. \\
I: The Physical Scenario} 
\author{S. Liberati${}^{\dagger}$}
\address{International School for Advanced Studies, Via Beirut 2-4, 
34014 Trieste, Italy\\
INFN sezione di Trieste}
\author{Matt Visser${}^{\P}$}
\address{Physics Department, Washington University, 
Saint Louis, Missouri 63130-4899, USA}
\author{F. Belgiorno${}^{\ddagger}$}
\address{\Universita\  degli Studi di Milano, Dipartimento di Fisica, 
Via Celoria 16,  20133 Milano, Italy}
\author{D.W. Sciama${}^{\S}$}
\address{International School for Advanced Studies, Via Beirut 2-4, 
34014 Trieste, Italy\\
International Center for Theoretical Physics,  Strada Costiera 11, 
34014 Trieste, Italy\\
Physics Department, Oxford University, Oxford, England\\
\hfill}

\date{5 April 1999; Typos fixed 10 May 1999; \LaTeX-ed \today}

\maketitle{}
\begin{abstract}
Several years ago Schwinger proposed a physical mechanism for
sonoluminescence in terms of changes in the properties of the
quantum-electrodynamic (QED) vacuum state during collapse of the
bubble.  This mechanism is most often phrased in terms of changes in
the Casimir Energy ({\em i.e.}, changes in the distribution of
zero-point energies) and has recently been the subject of considerable
controversy. The present paper further develops this quantum-vacuum
approach to sonoluminescence: We calculate Bogolubov coefficients
relating the QED vacuum states in the presence of a homogeneous medium
of changing dielectric constant. In this way we derive an estimate for
the spectrum, number of photons, and total energy emitted.  We
emphasize the importance of rapid spatio-temporal changes in
refractive indices, and the delicate sensitivity of the emitted
radiation to the precise dependence of the refractive index as a
function of wavenumber, pressure, temperature, and noble gas
admixture.  Although the basic physics of the dynamical Casimir effect
is a {\em universal\,} phenomenon of QED, specific and particular
experimental features are encoded in the condensed matter physics
controlling the details of the refractive index.  This calculation
places rather tight constraints on the possibility of using the
dynamical Casimir effect as an explanation for sonoluminescence, and
we are hopeful that this scenario will soon be amenable to direct
experimental probes. In a companion paper we discuss the technical
complications due to finite-size effects, but for reasons of clarity
in this paper we confine attention to bulk effects.
\end{abstract}


\section{Introduction}
\def\ng{n_{\mathrm gas}}
\def\ngi{n_{\mathrm gas}^{\mathrm in}}
\def\ngo{n_{\mathrm gas}^{\mathrm out}}
\def\nl{n_{\mathrm liquid}}
\def\nli{n_{\mathrm liquid}^{\mathrm in}}
\def\nlo{n_{\mathrm liquid}^{\mathrm out}}
\def\ni{n_{\mathrm in}}
\def\no{n_{\mathrm out}}
\def\nis{n_{\mathrm inside}}
\def\nos{n_{\mathrm outside}}
\def\Ni{ {\cal N}_{\mathrm in}}
\def\No{ {\cal N}_{\mathrm out}}
\def\max{\hbox{max}}
\def\min{\hbox{min}}
\def\in{{\mathrm in}}
\def\out{{\mathrm out}}
\def\sinc{\mathop{\hbox{sinc}}}
\def\half{{\textstyle{1\over2}}}
\def\quarter{{\textstyle{1\over4}}}
\def\omegai{\omega_\in}
\def\omegao{\omega_\out}
\def\observed{ {\mathrm observed} }

Sonoluminescence (SL) is the phenomenon of light emission by a
sound-driven gas bubble in fluid \cite{Physics-Reports}. In SL
experiments the intensity of a standing sound wave is increased until
the pulsations of a bubble of gas trapped at a velocity node have
sufficient amplitude to emit brief flashes of light having a
``quasi-thermal'' spectrum with a ``temperature'' of several tens of
thousands of Kelvin.  The basic mechanism of light production in this
phenomenon is still highly controversial. We first present a brief
summary of the main experimental data (as currently understood) and
their sensitivities to external and internal conditions.  For a more
detailed discussion see~\cite{Physics-Reports}.

\begin{figure}[htb]
\vbox{\hfil\psfig{figure=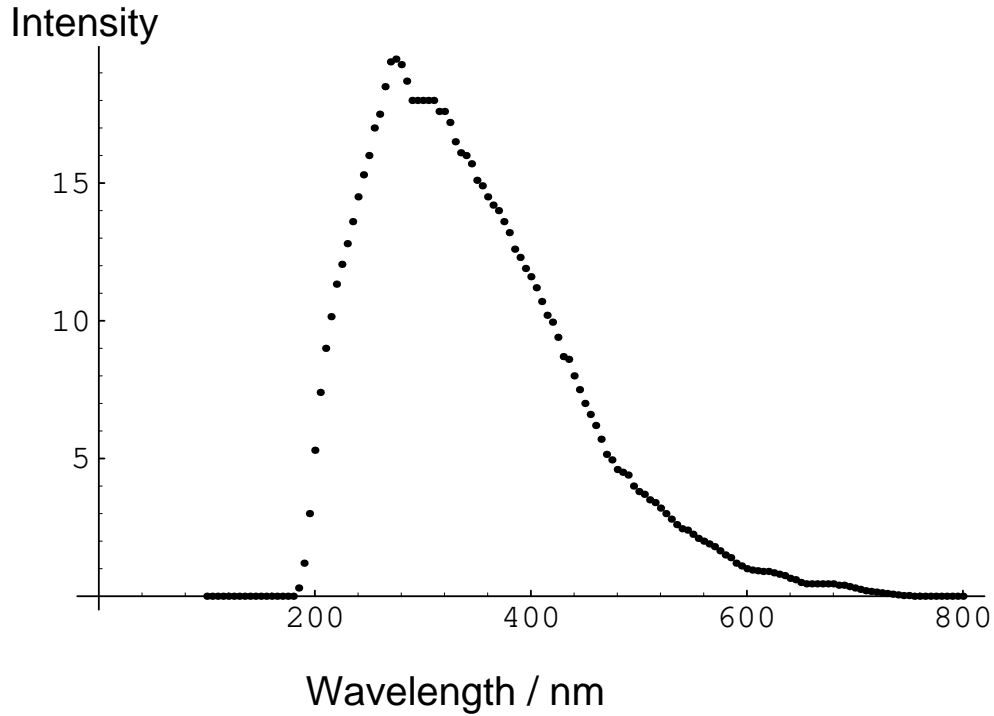,height=10cm}\hfil}
\caption{%
Typical experimental spectrum: The data has been extracted from
figure 51 of reference 1, and has here been plotted as intensity
(arbitrary units) as a function of wavelength. Note that no data
has been taken at frequencies below the visible range. The spectrum
is a broad-band spectrum without significant structure. The physical
nature of the cutoff (which occurs in the far ultraviolet) is one
of the key issues under debate.}
\label{F:experiment-1}
\end{figure}

\begin{figure}[htb]
\vbox{\hfil\psfig{figure=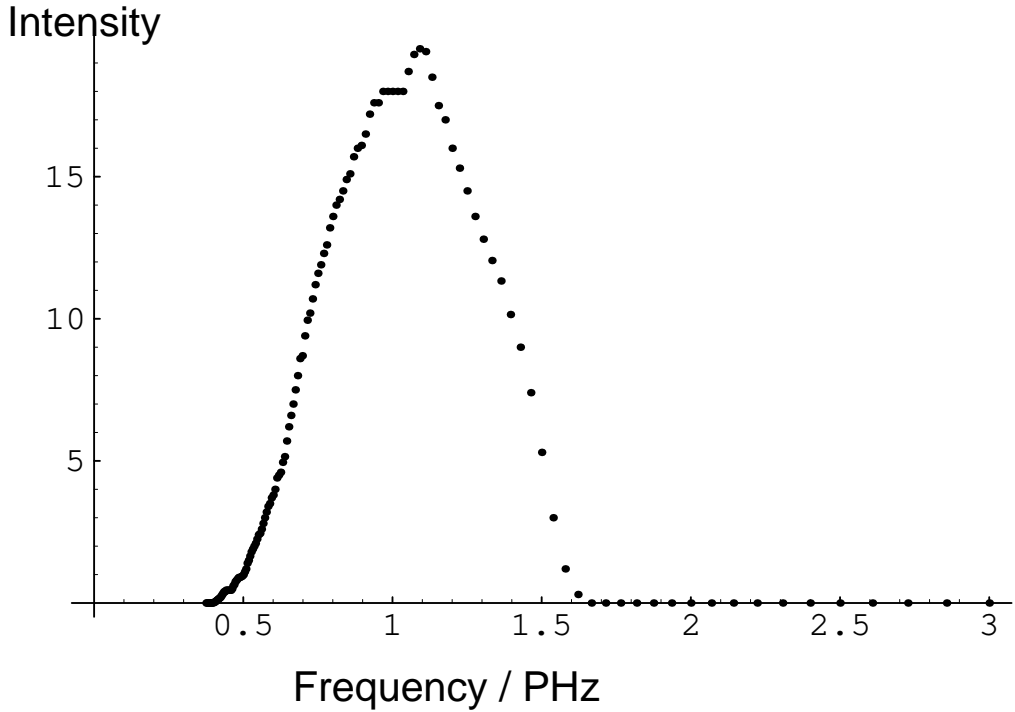,height=10cm}\hfil}
\caption{%
Typical experimental spectrum: The data has been extracted from
figure 51 of reference 1, and has here been plotted as
a number spectrum as a function of frequency.} 
\label{F:experiment-2} 
\end{figure}

SL experiments usually deal with bubbles of air in water, with
ambient radius $R_{\mathrm ambient} \approx 4.5 \; \mu {\rm m}$.
The bubble is driven by a sound wave of frequency of 20--30 kHz.
(Audible frequencies can also be used, at the cost of inducing
deafness in the experimental staff.)  During the expansion phase,
the bubble radius reaches a maximum of order $R_{\mathrm max}\approx
\; 45 \;\mu {\rm m}$, followed by a rapid collapse down to a minimum
radius of order $R_{\mathrm min}\approx 0.5 \; \mu {\rm m}$. The
photons emitted by such a pulsating bubble have typical wavelengths of
the order of visible light. The minimum observed wavelengths range
between $200\; {\rm nm}$ and $100\; {\rm nm}$. This light appears
distributed with a broad-band spectrum. (No resonance lines, roughly a
power-law spectrum with exponent depending on the noble gas admixture
entrained in the bubble, and with a cutoff in the extreme
ultraviolet.) For a typical example, see figures~\ref{F:experiment-1}
and~\ref{F:experiment-2}. If one fits the data to a Planck black-body
spectrum the corresponding temperature is several tens of thousands of
Kelvin (typically $70,000\; {\rm K}$, though estimates varying from
$40,000 \; {\rm K}$ to $100,000 \; {\rm K}$ are common). There is
considerable doubt as to whether or not this temperature parameter
corresponds to any real physical temperature.  There are about one
million photons emitted per flash, and the time-averaged total power
emitted is between $30$ and $100\;
\hbox{ mW}$.

The photons appear to be emitted by a very tiny spatio-temporal
region: Estimated flash widths vary from less than $35 \; {\rm ps}$ to
more than $380 \; {\rm ps}$ depending on the gas entrained in the
bubble \cite{Flash1,Flash2}. There are model-dependent (and
controversial) claims that the emission times and flash widths do not
depend on wavelength \cite{Flash2}. As for the spatial scale, there
are various model-dependent estimates but no direct measurement is
available \cite{Flash2}.  Though it is clear that there is a frequency
cutoff at about $1 \; {\rm PHz}$ the physics behind this cutoff is
controversial. Standard explanations are (1) a Thermal cutoff
(deprecated because the observed cutoff is much sharper than
exponential) , or (2) the opacity of water in the UV (deprecated
because of the observed absence of dissociation effects).
Alternatively, Schwinger suggests that the critical issue is that (3)
the real part of the refractive index of water goes to unity in the UV
(so that there is no change in the Casimir energy during bubble
collapse). We shall add another possible contribution to the mix: (4)
a rapidly changing refractive index causes photon production with an
``adiabatic cutoff' that depends on the timescale over which the
refractive index changes. (Because the observed falloff above the
physical cutoff is super-exponential it is clear that this adiabatic
effect is at most part of the complete picture.)

Any truly successful theory of SL must also explain a whole series of
characteristic sensitivities to different external and internal
conditions. Among these dependencies the main one is surely the
mysterious catalytic role of noble gas admixtures.  (Most often a few
percent in the air entrained in the bubble. One can obtain SL from air
bubbles with a $1\%$ content of argon, and also from pure noble gas
bubbles, but the phenomenon is practically absent in pure oxygen
bubbles.) In fact, it has been suggested that physical processes
concentrate the noble gasses inside the bubble to the extent that the
bubble consists of almost pure noble gas~\cite{Argon}, and some
experimental results seem to corroborate this
suggestion~\cite{Matula}.

Other external conditions that influence SL are: (1) Magnetic Fields
--- If the frequency of the driving sound wave is kept fixed, SL
disappears above a pressure-dependent threshold magnetic field: $H
\geq H_{0}(p)$. On the other hand, for a fixed value of the magnetic
field $H_{0}$, there are both upper and lower bounds on the applied
pressure that bracket the region of SL, and these bounds are
increasing functions of the applied magnetic field~\cite{YSK}.  This
is often interpreted as suggesting that the primary effect of magnetic
fields is to alter the condition for stable bubble
oscillations. See. also~\cite{Magnetic}. (2) Temperature of the water
--- If $T_{H_{2}O}$ decreases then the emitted power $W$ increases.
The position of the peak of the spectrum depends on $T_{H_{2}O}$. It
has been suggested that the increased light emission at lower water
temperature is associated with an increased stability of the bubble,
allowing for higher driving pressures~\cite{Hilgen}.

These are only the most salient features of the SL phenomenon.  In
attempting to explain such detailed and specific behaviour the
dynamical Casimir approach (QED vacuum approach) encounters the same
problems as all other approaches have. Nevertheless we shall argue
that SL explanations using a Casimir-like framework are viable, and
merit further investigation.

In this paper we shall concentrate on changes in the QED vacuum state
as a candidate explanation for SL, and try to clear up considerable
confusion as to what models based on the Casimir effect do and do not
predict. It is important to realize that changes in the static Casimir
energy in this experimental situation are big, that they have roughly
the right energy budget to drive SL, and that any purported
non-Casimir explanation for SL will have to find a way to hide the
effects of these changes in Casimir energy so as to make them
unobservable.

\section{Quantum-electrodynamic models of SL}
\subsection{Quasi-static Casimir models: Schwinger's approach}

The idea of a ``Casimir route'' to SL is due to Schwinger who several
years ago wrote a series of papers~\cite{Sc1,Sc2,Sc3,Sc4,Sc5,Sc6,Sc7}
regarding the so-called dynamical Casimir effect. Considerable
confusion has been caused by Schwinger's choice of the phrase
``dynamical Casimir effect'' to describe his particular model. In
fact, Schwinger's original model is not dynamical but is instead
quasi-static as the heart of the model lies in comparing two static
Casimir energy calculations: That for an expanded bubble with that for
a collapsed bubble. One key issue in Schwinger's model is thus simply
that of calculating Casimir energies for dielectric spheres---and
there is already considerable disagreement on this issue. A second and
in some ways more critical question is the extent to which this
difference in Casimir energies may be converted to real photons during
the collapse of the bubble---it is this issue that we shall address in
this paper.  The original quasi-static incarnation of the Schwinger
model had no real way of estimating either photon production
efficiency or timing information ({\em when} does the flash occur?).
In contrast the model of Eberlein~\cite{Eberlein1,Eberlein2,Eberlein3}
(more fully discussed below) is truly dynamical but uses a much more
specific physical approximation---the adiabatic approximation.  The
two models should not be confused. In this paper we shall argue that
the observed features of SL force one to make the sudden
approximation. We can then estimate the spectrum of the emitted
photons by calculating an appropriate Bogolubov coefficient relating
two states of the QED vacuum. The resulting variant of the Schwinger
model for SL is then rather tightly constrained, and should be
amenable to experimental verification (or falsification) in the near
future.

In his series of papers~\cite{Sc1,Sc2,Sc3,Sc4,Sc5,Sc6,Sc7} on SL,
Schwinger showed that the dominant bulk contribution to the Casimir
energy of a bubble (of dielectric constant $\epsilon_{\mathrm
inside}$) in a dielectric background (of dielectric constant
$\epsilon_{\mathrm outside}$) is~\cite{Sc4}
\begin{eqnarray}
E_{\mathrm cavity}
&=&
+2\frac{4\pi}{3}R^3 \; \int_0^K
{4 \pi k^2 dk\over(2\pi)^3}    \; \frac{1}{2}  \hbar c k
\left(
{1\over\sqrt{\epsilon_{\mathrm inside}}}- 
{1\over\sqrt{\epsilon_{\mathrm outside}}}
\right) +\cdots
\nonumber\\
&=&+\frac{1}{6 \pi} \hbar c R^3 K^4
\left(
{1\over\sqrt{\epsilon_{\mathrm inside}}}- 
{1\over\sqrt{\epsilon_{\mathrm outside}}}
\right) +\cdots.
\label{E:schwinger}
\end{eqnarray}
The corresponding number of emitted photons is
\begin{eqnarray}
N
&=&
+2\frac{4\pi}{3}R^3 \; \int_0^K
{4 \pi k^2 dk\over(2\pi)^3}    \; \frac{1}{2}
\left(
\frac{\sqrt{\epsilon_{\mathrm outside}}}{\sqrt{\epsilon_{\mathrm 
inside}}}- 1 \right) +\cdots
\nonumber\\
&=&+\frac{2}{9 \pi} (RK)^3
\left(
\frac{\sqrt{\epsilon_{\mathrm outside}}}{\sqrt{\epsilon_{\mathrm 
inside}}}- 1 \right) +\cdots.
\label{N:schwinger}
\end{eqnarray}   
Here we have inserted an explicit factor of two with respect to
Schwinger's papers to take into account both photon polarizations.
There are additional sub-dominant finite volume effects discussed
in~\cite{CMMV1,CMMV2,MV}. Schwinger's result can also be rephrased in
the clearer and more general form as~\cite{CMMV1,CMMV2,MV}:
\begin{equation}
E_{\mathrm cavity} = + 2 V \int \frac{d^3\vec{k}}{(2 \pi)^3} \; 
\frac{1}{2}
\hbar  \left[ \omega_{\mathrm inside}(k) - 
\omega_{\mathrm outside}(k)  \right] + \cdots
\end{equation}
\begin{equation}
N = + 2 V \int \frac{d^3\vec{k}}{(2 \pi)^3} \; 
\frac{1}{2}
 \left[ 
{\omega_{\mathrm inside}(k) \over \omega_{\mathrm outside}(k)} 
- 1  \right] + \cdots
\end{equation}
Here it is evident that the Casimir energy can be interpreted as a
difference in zero point energies due to the different dispersion
relations inside and outside the bubble.  The quantity $K$ appearing
above is a high-wavenumber cutoff that characterizes the wavenumber at
which the (real part of) the refractive indices ($n=\sqrt{\epsilon}$)
drop to their vacuum values. It is important to stress that this
cutoff it is not a regularization artifact to be removed at the end of
the calculation. The cutoff has a true physical meaning in its own
right.

The three main points of strength of models based on zero point
fluctuations ({\em e.g.}, Schwinger's model and its variants) are:

1) One does {\em not} need to achieve ``real'' temperatures of
thousands of Kelvin inside the bubble. As discussed in~\cite{2gamma},
quasi-thermal behaviour is generated in quantum vacuum models by the
squeezed nature of the two photon states created \cite{Eberlein2}, and
the ``temperature'' parameter is a measure of the squeezing, not a
measure of any real physical temperature\footnote{%
This ``false thermality'' must not be confused with the very specific
phenomenon of Unruh temperature.  In that case, valid only for
uniformly accelerated observers in flat spacetime, the temperature is
proportional to the constant value of the acceleration. Instead, in
the case of squeezed states, the apparent temperature can be related
to the degree of squeezing of the real photon pairs generated via the
dynamical Casimir effect.}.
(Of course, one should remember that the experimental data merely
indicates an approximately power-law spectrum [$N(\omega) \propto
\omega^\alpha$] with some sort of cutoff in the ultraviolet, and with
an exponent that depends on the gases entrained in the bubble; the
much-quoted ``temperature'' of the SL radiation is merely an
indication of the scale of this cutoff $K$.)

2) There is no actual production of far ultraviolet photons (because
the refractive index goes to unity in the far ultraviolet) so one does
not expect dissociation effects in water that other models
imply. Models based on the quantum vacuum automatically provide a
cutoff in the far ultraviolet from the behaviour of the refractive
index---this observation going back to Schwinger's first papers on the
subject.

3) One naturally gets the right energy budget. For $n_{\mathrm
outside} \approx 1.3$, $n_{\mathrm inside} \approx 1$, $K$ in the
ultra-violet, and $R\approx R_{\mathrm max}$, the change in the static
Casimir energy approximately equals the energy emitted each cycle.

This last point is still the object of heated debate.
Milton~\cite{M95}, and Milton and Ng~\cite{M96,M97} strongly criticize
Schwinger's result claiming that actually the Casimir energy contains
at best a surface term, the bulk term being discarded via (what is to
our minds) a physically dubious renormalization argument.  In their
more recent paper~\cite{M97} they discard even the surface term and
now claim that the Casimir energy for a dielectric bubble is of order
$E\approx \hbar c/R$.  (The dispute is ongoing---see~\cite{M98}.)
These points have been discussed extensively in~\cite{CMMV1,CMMV2,MV}
where it is emphasized that one has to compare two different physical
configurations of the same system, corresponding to two different
geometrical configurations of matter, and thus must compare {\em two}
different quantum states defined on the {\em same}
spacetime\footnote{%
This point of view is also in agreement with the bag model results of
Candelas~\cite{Candelas}.  It is easy to see that in the bag model one
finds a bulk contribution that happens to be zero only because of the
particular condition that $\epsilon \mu = 1$ everywhere. This
condition ensures the constancy of the speed of light (and so the
invariance of the dispersion relation) on all space while allowing the
dielectric constant to be less than one outside the vacuum bag (as the
model for quark confinement requires).}.
In a situation like Schwinger's model for SL one has to subtract from
the zero point energy (ZPE) for a vacuum bubble in water the ZPE for
water filling all space.  It is clear that in this case the bulk term
is physical and {\em must} be taken into account. Surface terms are
also present, and eventually other higher order correction terms, but
they prove to not be dominant for sufficiently large
cavities~\cite{MV}.

The calculations of Brevik {\em et al}~\cite{Brevik} and Nesterenko
and Pirozhenko~\cite{Nesterenko} also fail to retain the known bulk
volume term. In this case the subtlety in the calculation arises from
neglecting the continuum part of the spectrum. They consider a
dielectric sphere in an infinite dielectric background and sum {\em
only} over the discrete part of the spectrum to calculate their
Casimir ``energy''. When the continuum modes are reintroduced the
proper volume dependence is recovered.  Their conclusions regarding
the relevance of the Casimir effect to SL are then incorrect: by
completely discarding the volume (and indeed surface) contribution
they are left with a Casimir ``energy'' that can be simply estimated
by dimensional analysis to be of order $\hbar c/R$ and is strictly
proportional to the inverse radius of the bubble.  This is certainly a
very small quantity insufficient to drive SL but this is also not the
correct physical quantity to calculate.  For a careful discussion of
the correct identification of the physically relevant Casimir energy
see~\cite{CMMV1,CMMV2,MV}.

While we believe that the contentious issues of how to define the
Casimir energy are successfully dealt with in~\cite{CMMV1,CMMV2,MV},
one of the subsidiary aims of this paper is to side-step this whole
argument and provide an independent calculation demonstrating
efficient photon production.

In contrast to the points of strength outlined above, the main
weakness of the original quasi-static version of Schwinger's idea is
that there is no real way to calculate either timing information or
conversion efficiency.  A naive estimate is to simply and directly
link power produced to the change in volume of the bubble.  As pointed
out by Barber {\em et al.}~\cite{Physics-Reports}, this assumption
would imply that the main production of photons may be expected when
the the rate of change of the volume is maximum, which is
experimentally found to occur near the maximum radius. In contrast the
emission of light is experimentally found to occur near the point of
minimum radius, where the rate of change of area is maximum. All else
being equal, this would seem to indicate a surface dependence and
might be interpreted as a true weakness of the dynamical Casimir
explanation of SL.  In fact we shall show that the situation is
considerably more complex than might naively be thought. We claim that
there is much more going on than a simple change in volume of the
bubble, and shall shortly focus attention on the ``bounce'' that
occurs as the contents of the bubble hit the van der Waals hard core
maximum density.

\subsection{Eberlein's dynamical model for SL}

The quantum vacuum approach to SL was developed extensively in the
work of Eberlein~\cite{Eberlein1,Eberlein2,Eberlein3}.  The basic
mechanism in Eberlein's approach is a particular implementation of the
dynamical Casimir effect: Photons are assumed to be produced due to an
{\em adiabatic} change of the refractive index in the region of space
between the minimum and the maximum bubble radius (a related
discussion for time-varying but spatially-constant refractive index
can be found in the discussion by Yablonovitch\cite{Yablonovitch}).
This physical framework is actually implemented via a boundary between
two dielectric media which accelerates with respect to the rest frame
of the quantum vacuum state. The change in the zero-point modes of the
fields produces a non-zero radiation flux.  Eberlein's contribution
was to take the general phenomenon of photon generation by moving
dielectric boundaries and attempt a specific implementation of these
ideas as a candidate for explaining SL.

It is important to realize that this is a second-order effect.  Though
he was unable to provide a calculation to demonstrate it, Schwinger's
original discussion is posited on the direct conversion of zero point
fluctuations in the expanded bubble vacuum state into real photons
plus zero point fluctuations in the collapsed bubble vacuum
state. Eberlein's mechanism is a more subtle (and much weaker) effect
involving the response of the atoms in the dielectric medium to
acceleration through the zero-point fluctuations. The two mechanisms
are quite distinct and considerable confusion has been engendered by
conflating the two mechanisms. Criticisms of the Eberlein mechanism do
not necessarily apply to the Schwinger mechanism, and vice versa.

In the Eberlein analysis the motion of the bubble boundary is taken
into account by introducing a velocity-dependent perturbation to the
usual EM Hamiltonian:
\begin{eqnarray}
H_{\epsilon} &\!=&\! 
\frac{1}{2} \int{\rm d}^3{\bf r} 
\left(
{{\bf D}^2\over\epsilon} + {\bf B}^2 
\right)\;,
\\
\Delta H &\!=&\!  
\beta
\int{\rm d}^3{\bf r} 
\frac{\epsilon -1}{\epsilon}\;
({\bf D}\wedge{\bf B})\cdot{\bf\hat r}\;.
\end{eqnarray}
This is an approximate low-velocity result coming from a power series
expansion in the speed of the bubble wall $\beta= \dot R/c$.  The
bubble wall is known to collapse with supersonic velocity, values of
Mach 4 are often quoted, but this is still completely non-relativistic
with $\beta\approx 10^{-5}$. Unfortunately, when the Eberlein
formalism is used to model the observed quantity of radiation from
each SL flash the implied bubble wall velocities are superluminal,
indicating that one has moved outside the region of validity of the
approximation scheme~\cite{Eberlein-discussion}.

The Eberlein approach consists of a novel mixture of the standard
adiabatic approximation with perturbation theory. In principle, the
adiabatic approximation requires the knowledge of the complete set of
eigenfunctions of the Hamiltonian for any allowed value of the
parameter. In the present case only the eigenfunctions of part of the
Hamiltonian, namely those of $H_\epsilon$, are known. (And these
eigenfunctions are known explicitly only in the adiabatic
approximation where $\epsilon$ is treated as time-independent.)  The
calculation consists of initially invoking the standard application of
the adiabatic approximation to the full Hamiltonian, then formally
calculating the transition coefficients for the vacuum to two photon
transition to first order in $\beta$, and finally in explicitly
calculating the radiated energy and spectral density. In this last
step Eberlein used an approximation valid only in the limit $k R\gg1$
which means in the limit of photon wavelengths smaller than the bubble
radius. (Compare this with our discussion of the large $R$ limit
below.) This implies that the calculation will completely miss any
resonances that are present.

Eberlein's final result for the energy radiated over one acoustic
cycle is:
\begin{equation}
{\cal W} = 1.16\:\frac{(n^2-1)^2}{n^2}\,\frac{1}{480\pi}
\left[{\hbar\over c^3}\right]
\int_{0}^{T} {\rm d}\tau\; \frac{\partial^5 R^2(\tau)}{\partial
        \tau^5}\,R(\tau)\beta(\tau)\;.
\end{equation}
Eberlein approximates $n_{\mathrm inside} \approx n_{\mathrm air}
\approx 1$ and sets $n_{\mathrm outside} = n_{\mathrm water} \to
n$. The $1.16$ is the result of an integration that has to be
estimated numerically. The precise nature of the semi-analytic
approximations made as prelude to performing the numerical integration
are far from clear.

One of the interesting consequences of this result is that the
dissipative force acting on the moving dielectric interface can be
seen to behave like $R^2\beta^{(4)}(t)$, plus terms with lower
derivatives of $\beta$. This dependence tallies with results of
calculations for frictional forces on moving perfect mirrors; the
dissipative part of the radiation pressure on a moving dielectric
{\em half-space} or {\em flat} mirror is proportional to the fourth
derivative of the velocity~\cite{JR}.

By a double integration by parts the above can be re-cast as
\begin{equation}
{\cal W} = 1.16\:\frac{(n^2-1)^2}{n^2}\,\frac{1}{960\pi}
\left[{\hbar\over c^4}\right]
\int_{0}^{T} {\rm d}\tau\; 
\left(\frac{\partial^3 R^2(\tau)}{\partial \tau^3}\right)^2.
\end{equation}
Then the energy radiated is also seen to be proportional to
\begin{equation}
\int_0^T (\dot R)^2 (\ddot R)^2 + ...
\end{equation}
explicitly showing that the acceleration of the interface ($\ddot
R$) and the strength of the perturbation ($\dot R$) both contribute
to the radiated energy.

The main improvement of this model over the original Schwinger
model is the ability to provide basic timing information: In this
mechanism the massive burst of photons is produced at and near the
turn-around at the minimum radius of the bubble. There the velocity
rapidly changes sign, from collapse to re-expansion. This means
that the acceleration is peaked at this moment, and so are higher
derivatives of the velocity.  Other main points of strength of the
Eberlein model are the same as previously listed for the Schwinger
model. However, Eberlein's model exhibits significant weaknesses
(which do not apply to the Schwinger model):

1) The calculation is based on an adiabatic approximation which
does not seem consistent with results.  The adiabatic approximation
would seem to be justified in the SL case by the fact that the
frequency $\Omega$ of the driving sound is of the order of tens of
kHz, while that of the emitted light is of the order of $10^{15}$
Hz. But if you take a timescale for bubble collapse of the order
of milliseconds, or even microseconds, then photon production is
extremely inefficient, being exponentially suppressed [as we shall
soon see] by a factor of $\exp(-\omega/\Omega)$.

In order to compare with the experimental data the model requires,
as external input, the time dependence of the bubble radius.  This
is expressed as a function of a parameter $\gamma$ which describes
the time scale of the collapse and re-expansion process.  In order
to be compatible with the experimental values for emitted power
$\cal W$ one has to fix $\gamma \approx 10\; {\rm fs}$. This is
far too short a time to be compatible with the {\em adiabatic}
approximation.  Although one may claim that {\em the precise
numerical value of the timescale} can ultimately be modified by
the eventual inclusion of resonances it would seem reasonable
to take this ten femtosecond figure as a first self-consistent
approximation for the characteristic timescale of the driving system
(the pulsating bubble).  Unfortunately, the characteristic timescale
of the collapsing bubble then comes out to be of the same order of
the characteristic period of the emitted photons, {\em violating
the adiabatic approximation used in deriving the result}. Attempts
at bootstrapping the calculation into self-consistency instead
bring it to a regime where the adiabatic approximation underlying
the scheme cannot be trusted.

2) The Eberlein calculation  cannot deal with any resonances that
may be present.  Eberlein does consider resonances to be a possible
important correction to her model, but she is considering ``classical''
resonances (scale of the cavity of the same order of the wavelength
of the photons) instead of what we feel is the more interesting
possibility of parametric resonances.

Finally we should mention a recent calculation that gives
qualitatively the same results as the Eberlein model although leading
to different formulae. \Schutzhold, Plunien, and Soff~\cite{SPS} adopt
a slightly different decomposition into unperturbed and perturbing
Hamiltonians by taking
\begin{eqnarray}
H_{0} &\!=&\! 
\frac{1}{2} \int{\rm d}^3{\bf r} 
\left(
{{\bf D}^2\over \epsilon_0} + {\bf B}^2 
\right),
\\
\Delta H &\!=&\!  
\int{\rm d}^3{\bf r} 
\left(
-\frac{1}{2} {\epsilon-\epsilon_0\over\epsilon\epsilon_0} {\bf D}^2 
+ {\bf \beta} \; 
\frac{\epsilon -\epsilon_0}{\epsilon}\;
({\bf D}\wedge{\bf B}) \cdot {\rm \hat r}
\right),
\end{eqnarray}
Their result for the total energy emitted per cycle is given
analytically by
\begin{equation}
{\cal W} = {n^2(n^2-1)^2 \over 1890\pi}
\left[{\hbar\over c^6}\right]
\int_{0}^{T} {\rm d}\tau\; 
\left(\frac{\partial^4 R^3(\tau)}{\partial \tau^4}\right)^2.
\end{equation}
The key differences are that this formula is analytic (rather than
numerical) and involves fourth derivatives of the volume of the
bubble (rather than third derivatives of the surface area).  The
main reason for the discrepancy between this and Eberlein's result
can be seen as due to a different choice of the dependence on $r$
of $\beta(r,t)$. In reference~\cite{SPS} they considered the more
physical case of a localized disturbance that yields significant
contributions only over a bounded volume. (Eberlein makes the
simplifying assumption that $\beta(r,t)$ is a function of $t$ only,
which is incompatible with continuity and the essentially constant
density of water. In contrast, \Schutzhold\ {\em et al.} take the
radial velocity of the water outside the bubble to be $\beta(r,t)
= f(t)/r^2$.)

Putting these models aside, and before proposing new routes for
developing further research in SL, we shall give below a more detailed
discussion of some important points of Schwinger's model which seem to
us to be crucial in order to understand the possibility of a vacuum
explanation of SL.

\subsection{Timescales: The need for a sudden approximation.}

One of the key features of photon production by a space-dependent and
time-dependent refractive index is that for a change occurring on a
timescale $\tau$, the amount of photon production is exponentially
suppressed by an amount $\exp(-\omega\tau)$. Below we provide a
specific model that exhibits this behaviour, and argue that the result
is in fact generic.

The importance for SL is that the experimental spectrum is {\em not\,}
exponentially suppressed at least out to the far ultraviolet.
Therefore any mechanism of Casimir-induced photon production based on
an adiabatic approximation is destined to failure: Since the
exponential suppression is not visible out to $\omega \approx 10^{15}
\hbox{ Hz}$, it follows that {\em if\,} SL is to be attributed to
photon production from a time-dependent refractive index ({\em i.e.},
the dynamical Casimir effect), {\em then} the timescale for change in
the refractive index must be of order a {\em femtosecond\/}\footnote{
Actually, once one takes into account the refractive index of the
final state this condition can be somewhat relaxed. We ultimately find
that we can tolerate a refractive index that changes as slowly as on a
picosecond timescale, but this is still far to rapid to be associated
with physical collapse of the bubble. For the time being we focus on
the femtosecond timescale (which actually makes things more difficult
for us) to check the physical plausibility of the scenario, but keep
in mind that eventually things can be relaxed by a few orders of
magnitude.}.
Thus any Casimir--based model has to take into account that {\em
the change in the refractive index cannot be due just to the change
in the bubble radius}.

The SL flash is known to occur at or shortly after the point of
maximum compression. The light flash is emitted when the bubble is at
or near minimum radius $R_{\mathrm min} \approx 0.5\;\mu \hbox{m} =
500\;\hbox{nm}$.  Note that to get an order femtosecond change in
refractive index over a distance of about $500\;{\rm nm}$, the change
in refractive index has to propagate at relativistic speeds.  To
achieve this, we must adjust basic aspects of the model: We will move
away from the original Schwinger suggestion, in that it is no longer
the collapse from $R_{\mathrm max}$ to $R_{\mathrm min}$ that is
important.  {\em Instead we will postulate a rapid (order femtosecond)
change in refractive index of the gas bubble when it hits the van der
Waals hard core.}

The underlying idea is that there is some physical process that gives
rise to a sudden change of the refractive index inside the bubble when
it reaches maximum compression.  We have to ensure that the velocity
of mechanical perturbations, that is the sound velocity, can be a
significant fraction of the speed of light in this critical
regime\footnote{
This condition can also be slightly relaxed: One can conceive of the
change in refractive index being driven by a shockwave that appears at
the van der Waals hard core. Now a shockwave is by definition a
supersonic phenomenon. If the velocity of sound is itself already
extremely high then the shockwave velocity may be even higher.  Note
that most of the viable models for the gas dynamics during the
collapse predict the formation of strong shock-waves; so we are
adapting physics already envisaged in the literature. At the same time
we are asking for less extreme conditions ({\em e.g.}, we can be much
more relaxed regarding the focussing of these shockwaves) in that we
just need a rapid change of the refractive index of the entrained gas,
and do not need to propose any overheating to ``stellar''
temperatures.  It is also interesting to note that changes in the
refractive index (but that of the surrounding water) due to the huge
compression generated by shockwaves have already been considered in
the literature~\cite{HRB} ({\em cf.} page 5437).}.
We first show that the minimum radius experimentally observed is of
the same order as the van der Waals hard core radius $R_{\mathrm
hc}$. The latter can be deduced as follows: It is known that the van
der Waals excluded volume for air is $b=$0.036
l/mol~\cite{Wu-Roberts}.  The minimum possible value of the volume is
then $V_{\mathrm hc} = b\cdot (\rho V_{\mathrm ambient})/m$, where
$(\rho V_{\mathrm ambient})/m$ gives the number of moles and
$V_{\mathrm ambient}$ is the ambient value of the volume. {From}
$R_{\mathrm hc} = R_{\mathrm ambient} \cdot (b
\rho/\mu)^{1/3}$ and assuming for the density of air $\rho=10^{-3}\;
{\rm gr}/{\rm cm}^3$ [$1.3 \times 10^{-3} \; {\rm gr}/{\rm cm}^3$ at
STP (standard temperature and pressure)], one gets $R_{\mathrm hc}\sim
0.48\; \mu {\mathrm m}$.  This value compares favorably with the
experimentally observed value of $R_{\mathrm min}$.  Moreover, the
role of the van der Waals hard core in limiting the collapse of the
bubble is suggested in~\cite{Physics-Reports} (cf. fig. 10, p.78), and
a careful hydrodynamic analysis for the case of an Argon
bubble~\cite{Fluid} reveals that for sonoluminescence it is necessary
that the bubble undergoes a so called strongly collapsing phase where
its minimum radius is indeed very near the hard core radius\footnote{
Noticeably, from~\cite{Fluid} it is easy to estimate the van der Waals
radius for the Argon bubble: $R_{\mathrm hc} \approx R_{\mathrm
ambient}/8.86$, with $R_{\mathrm ambient}=0.4 \mu {\rm m}$.  This
again gives $R_{\mathrm hc}(Argon) \approx 0.45 \; \mu {\rm m}
\approx R_{\mathrm min}$.}.

It is crucial to realize that a van der Waals gas, when compressed to
near its maximum density, has a speed of sound that goes relativistic.
To see this, write the (non-relativistic) van der Waals equation of
state as
\begin{equation}
p 
= {n k T \over 1-n b} - a n^2 
= {\rho kT/m \over 1- \rho/\rho_{\mathrm max}} 
- a {\rho^2\over m^2}.
\end{equation}
Here $n$ is the number density of molecules; $\rho$ is mass density;
$m$ is average molecular weight ($m=28.96\; \hbox{amu/molecule} =
28.96\; \hbox{gr/mol}$ for air).

Now consider the (isothermal) speed of sound for a van der Waals gas
\begin{equation}
v_{\mathrm sound}^2 
= \left({\partial p \over \partial \rho}\right)_{T} 
= {(kT/m) \over (1- \rho/\rho_{\mathrm max})^2} 
- 2a {\rho\over m^2}.
\end{equation}
Near maximum density this is
\begin{equation}
v_{\mathrm sound} 
\approx {\sqrt{kT/m} \over (1- \rho/\rho_{\mathrm max}) },
\end{equation}
and so it will go relativistic for densities close enough to maximum
density. It is {\em only} for the sake of simplicity that we have
considered the isothermal speed of sound. We do not expect the
process of bubble collapse and core bounce to be isothermal.
Nevertheless, this calculation is sufficient to demonstrate that
in general the sound velocity becomes formally infinite (and it is
reasonable that it goes relativistic) at the incompressibility
limit (where it hits the van der Waals hard core).  This conclusion
is not limited to the van der Waals equation of state, and is not
limited to isothermal (or even isentropic) sound propagation.
Indeed, consider any equation of state of the form
\begin{equation}
   p = \kappa({\cal V}-b,T),
\end{equation}
where ${\cal V}$ is the volume per mole occupied by the gas and 
$b$ is the ``minimum molar volume'', related to the molar mass and
maximum mass density by
\begin{equation}
   b = m/ \rho_{\mathrm max}.
\end{equation}

To say that $b$ is a minimum molar volume means that we want 
\begin{equation}
   \lim_{{\cal V} \to b} \kappa({\cal V}-b,T) = \infty.
\end{equation}
We can enforce this by demanding
\begin{equation}
   p({\cal V},T) = \kappa({\cal V}-b,T) = {\kappa_1({\cal V}-b,T)\over 
{\cal V}-b} + \kappa_2({\cal V}-b,T), 
\end{equation}
or equivalently
\begin{equation}
   p(\rho,T) = 
   {\kappa_3(\rho-\rho_{\mathrm max},T)\over
   1-\rho/\rho_{\mathrm max}} + \kappa_4(\rho-\rho_{\mathrm max},T),
\end{equation}
with the $\kappa_i({\cal V}-b,T)$ being less singular than $p({\cal
V}-b,T)$.  For typical model equations of state $\kappa_1({\cal
V}-b,T)$ and $\kappa_2({\cal V}-b,T)$ are typically differentiable and
finite.  (The van der Waals, Dieterici, Bethelot, and ``modified
adiabatic'' equations are all of this form, but the Moss {\em et al}
equation of state is {\em not} of this
form~\cite{Moss-et-al}.)~\footnote{%
The Dieterici equation of state is
\[
p = {n k T \over 1 - nb} \exp\left(-{a n\over k T}\right) 
  = {\rho k T /m \over 1 - \rho/\rho_{\mathrm max}} 
     \exp\left(-{a \rho\over m k T}\right),
\]
while the Bethelot equation of state is
\[
p = {n k T \over 1 - nb} - {a' n^2 \over T}
  = {\rho k T/m \over 1 -  \rho/\rho_{\mathrm max}} 
  - {a' \rho^2 \over m^2 T},
\]
so that it is a modified van der Waals equation with a particular
temperature dependence for the $a$ parameter ($a \to a'/T$).  The
``modified adiabatic'' equation of state discussed by Barber {\em
et al}~\cite{Physics-Reports} is
\[
p = p_0 \left( { 1 - n_0 b \over 1- n b} \right)^\gamma = 
    p_0 \left( { 1 - \rho_0/\rho_{\mathrm max} 
           \over 1 - \rho/\rho_{\mathrm max} } \right)^\gamma.
\]
In contrast, Moss et al~\cite{Moss-et-al} use a model equation of
state of the form
\[
p = {\rho k T\over m}(1 + \kappa) 
  + {\gamma E_c \rho \over 1-\gamma} 
    \left[ \left({\rho\over\rho_0}\right)^\gamma 
          - \left({\rho\over\rho_0}\right) 
    \right],
\]
with $\kappa$ and $\gamma$ being adjustable parameters.  This
equation of state does {\em not} exhibit a maximum hard-core
density.} 
Now calculate the speed of sound, keeping some quantity ``$X$''
constant
\begin{equation}
  v_{\mathrm sound}|_X 
      = \sqrt{ \left.\left({dp\over d\rho}\right)\right|_X }
      = \sqrt{ \left({\partial p\over\partial \rho}\right) 
             + \left({\partial p\over\partial T}\right) 
               \left.\left({dT\over d\rho}\right)\right|_X }.
\end{equation}
Then 
\begin{eqnarray}
 v_{\mathrm sound}^2|_X &=& 
        {\kappa_3(\rho-\rho_{\mathrm max},T)  \over 
         \rho_{\mathrm max} (1-\rho/\rho_{\mathrm max})^2 }
        + 
        \left({\partial \kappa_3\over\partial \rho}\right) 
        {1\over 1-\rho/\rho_{\mathrm max}}
        + 
        \left({\partial \kappa_3\over\partial T}\right) 
          \left.\left({dT\over d\rho}\right)\right|_X  
          {1\over 1-\rho/\rho_{\mathrm max}}
\nonumber\\
&&
\qquad
        + 
        \left({\partial \kappa_4\over\partial \rho}\right)
       + 
        \left({\partial \kappa_4\over\partial T}\right)
          \left.\left({dT\over d\rho}\right)\right|_X .
\end{eqnarray}
The net result is that as $v \to b$; {\em  i.e.} $ \rho \to
\rho_{\mathrm max}$; the speed of sound becomes relativistic
(formally infinite), independent of whether or not this is constant
temperature, constant entropy, or whatever ``constant $X$'' may
be\footnote{
It should be noted that similar unphysical features also affect the
``shock wave''--based models~\cite{Physics-Reports}: Indeed, the Mach
number of the shock formally diverges as the shock implodes towards
the origin ({\em cf.} page 126 of \cite{Physics-Reports}).  One way to
overcome this type of problem is the suggestion that that very near
the minimum radius of the bubble there is a breakdown of the
hydrodynamic description~\cite{Fluid}. If so, the thermodynamic
description in terms of state equations should probably be considered
to be on a heuristic footing at best.}.
There is something of a puzzle in the fact that hydrodynamic
simulations of bubble collapse do not see these relativistic effects.
Notably, the simulations by Moss {\em et al}~\cite{Moss-et-al} seem to
suggest collapse, shock wave production, and re-expansion all without
ever running into the van der Waals hard core. The fundamental reason
for this is that the model equation of state they choose does not have
a hard core for the bubble to bounce off\footnote{%
For low densities their equation of state is essentially perfect
gas, for medium densities it becomes stiffer (and the speed of
sound goes up), but for very high densities it again becomes softer,
and there is no ``maximum density''.  One might think that their
$\rho_0$, the density of solid air at zero Celsius, is a maximum
density, but if you look carefully at their equation of state the
gas is still compressible at this density; the pressure and speed
of sound are both finite.}.
Instead, there are a number of free parameters in their equation of
state which are chosen in such a way as to make their equation of
state stiff at intermediate densities, even if their equation of state
is by construction always soft at van der Waals hard core
densities. If the equation of state is made sufficiently stiff at
intermediate densities then a bounce can be forced to occur long
before hard core densities are encountered.  Unfortunately, as
previously mentioned, the experimental data and hydrodynamic analysis
seem to indicate that hard core densities {\em are} achieved at
maximum bubble compression.

Given all this, the use of a relativistic sound speed
is now physically justifiable, and the possibility of femtosecond
changes in the refractive index is at least physically plausible
(even though we cannot say that femtosecond changes in refractive
index are guaranteed).

So our new physical picture is this: The ``in state'' is a small
sphere of gas, radius about 500 nm, with some refractive index
$\ngi$ embedded in water of refractive index $\nl$. There is a
sudden femtosecond change in refractive index, essentially at
constant radius, so the ``out state'' is gas with refractive index
$\ngo$ embedded in water of refractive index $\nl\;$\footnote{%
In view of this femtosecond change of refractive index, we would
be justified in making the sudden approximation for frequencies
less than about a PHz.  In this paper we do not make this
approximation except when convenient in obtaining crude analytic
estimates, but when we turn to dealing with finite-size effects in
the companion paper~\cite{Companion} the sudden approximation will
be more than just a convenience: it will be absolutely essential
in keeping the mathematical features of the analysis tractable.}.
Thus our calculations will be complementary to Eberlein's calculations.
She was driven to femtosecond timescales to fit the experimental
data, but these femtosecond timescales then unfortunately undermined
the adiabatic approximation used to derive her results. In contrast,
we shall maintain a physically consistent calculation throughout.
These arguments have now pushed dynamical Casimir effect models
for SL into a rather constrained region of parameter space. We hope
that these ideas will become experimentally testable in the near
future.

\section{Bogolubov coefficients}

As a first approach to the problem of estimating the spectrum and
efficiency of photon production we decided to study in detail the
basic mechanism of particle creation and to test the consistency
of the Casimir energy proposals previously described.  With this
aim in mind we studied the effect of a changing dielectric constant
in a homogeneous medium.  At this stage of development, we are not
concerned with the detailed dynamics of the bubble surface, and
confine attention to the bulk effects, deferring consideration of
finite-volume effects to the companion paper~\cite{Companion}.

We shall consider two different asymptotic configurations.  An
``in'' configuration with refractive index $\ni$, and an ``out''
configuration with a refractive index $\no$.  These two configurations
will correspond to two different bases for the quantization of the
field.  (For the sake of simplicity we take, as Schwinger did, only
the electric part of QED, reducing the problem to a scalar
electrodynamics).  The two bases will be related by Bogolubov
coefficients in the usual way.  Once we determine these coefficients
we easily get the number of created particles per mode and from
this the spectrum.  Of course it is evident that such a model cannot
be considered a fully complete and satisfactory model for SL.  This
present calculation must still be viewed as a test calculation in
which basic features of the Casimir approach to SL are investigated.

In the original version of the Schwinger model it was usual to
simplify calculations by using the fact that the dielectric constant
of air is approximately equal 1 at standard temperature and pressure
(STP), and then dealing only with the dielectric constant of water
($n_{\mathrm liquid} = \sqrt{\epsilon_{\mathrm outside}} \approx
1.3$). We wish to avoid this temptation on the grounds that the
sonoluminescent flash is known to occur within $500$ picoseconds of
the bubble achieving minimum radius. Under these conditions the gases
trapped in the bubble are close to the absolute maximum density
implied by the hard core repulsion incorporated into the van der Waals
equation of state.  Gas densities are approximately one million times
atmospheric and conditions are nowhere near STP.  For this reason we
shall explicitly keep track of both initial and final refractive indices.

We now describe a simple analytically tractable model for the
conversion of zero point fluctuations (Casimir energy) into real
photons.  The model describes the effects of a time-dependent
refractive index in the infinite volume limit. We shall show that
for sudden changes in the refractive index the conversion of
zero-point fluctuations is highly efficient, being limited only by
phase space, whereas adiabatic changes of the refractive index lead
to exponentially suppressed photon production.

\def\curl{\nabla \times}
\def\div{\nabla \cdot}
\def\grad{\nabla}

\subsection{Defining the model}

Take an infinite homogeneous dielectric with a permittivity
$\epsilon(t)$ that depends only on time, not on space. The homogeneous
($dF=0$) Maxwell equations are
\begin{equation}
B = \curl A;
\end{equation}
\begin{equation}
E = - \grad \phi - {1\over c} {\partial A\over\partial t};
\end{equation}
while the source-free inhomogeneous ($*d*F=0$) Maxwell equations
become
\begin{equation}
\div (\epsilon E) = 0;
\end{equation}
\begin{equation}
\curl {B} = 
+{1\over c} {\partial\over\partial t} (\epsilon E). 
\end{equation}
Substituting into this last equation
\begin{equation}
\curl\left(\curl A\right)  =
- {1\over c} {\partial\over\partial t}
\left[  \epsilon \left( 
\grad \phi + {1\over c} {\partial A\over\partial t}
\right) \right].
\end{equation}
Suppose that $\epsilon(t)$ depends on time but not space, then
\begin{equation}
( \grad (\div A) - \nabla^2 A )= 
-  \grad {1\over c} {\partial\over\partial t}
(\epsilon \phi ) -  {1\over c^2 } {\partial\over\partial t} \epsilon 
{\partial A\over\partial t}.
\end{equation}
Now adopt a {\em generalized} Lorentz gauge
\begin{equation}
\div A  + 
{1\over c} {\partial\over\partial t} (\epsilon \phi ) = 0.
\end{equation}
Then the equations of motion reduce to 
\begin{equation}
{1\over c^2 } {\partial\over\partial t} \epsilon 
{\partial A\over\partial t}  = \nabla^2 A.
\end{equation}
We now introduce a ``pseudo-time'' parameter by defining
\begin{equation}
{\partial\over\partial \tau} = \epsilon(t) {\partial \over \partial t}.
\end{equation}
That is
\begin{equation}
\tau(t) = \int {dt\over\epsilon(t)}.
\end{equation}
In terms of this pseudo-time parameter the equation of motion is 
\begin{equation}
{\partial^2\over\partial\tau^2} A = 
c^2 {\epsilon(\tau)}\nabla^2 A.
\label{eqm}
\end{equation}
Compare this with equation (3.86) of Birrell and
Davies~\cite{Birrell-Davies}. Now pick a convenient profile for
the permittivity and permeability as a function of this pseudo-time.
(This particular choice of time profile for the refractive index
is only to make the problem analytically tractable, with a little
more work it is possible to consider generic monotonic changes of
refractive index and place bounds on the Bogolubov
coefficients~\cite{Visser}.) Let us take
\begin{eqnarray}
\label{E:profile}
{\epsilon(\tau)} &=& a + b \tanh(\tau/\tau_0)
\\
&=& \half (\ni^2 + \no^2) + \half(\no^2-\ni^2)\;\tanh(\tau/\tau_0).
\end{eqnarray}
Here $\tau_0$ represents the typical timescale of the change of the
refractive index in terms of the pseudo-time we have just defined.  We
are interested in computing the number of particles that can be
created passing from the ``in'' state ($t\to-\infty$, that is,
$\tau\to-\infty$) to the ``out'' state ($t\to+\infty$, that is,
$\tau\to+\infty$). This means we must determine the Bogolubov
coefficients that relate the ``in'' and ``out'' bases of the quantum
Hilbert space.  Defining the inner product as:
\begin{equation}
(\phi_1,\phi_2) =
i \int_{\Sigma_\tau} \phi_1^* \;
{\stackrel{\leftrightarrow}{\partial}\over\partial\tau}  \;
\phi_2\: d^3x,
\end{equation}  
The Bogolubov coefficients can now be {\em defined} as\\
\begin{eqnarray}
\alpha_{ij}
&=&
({A_{i}^\out },{A_{j}^\in }),
\\
\beta_{ij}
&=&(
{A_{i}^\out }^{*}, {A_{j}^\in }).
\end{eqnarray}
Where ${A_{i}^\in }$ and ${A_{j}^\out }$ are solutions
of the wave equation (\ref{eqm}) in the remote past and remote future
respectively.  We shall compute the coefficient $\beta_{ij}$ It is
this quantity that is linked to the spectrum of the ``out'' particles
present in the ``in'' vacuum, and it is this quantity that is related
to the total energy emitted.  With a few minor changes of notation we
can just write down the answers directly from pages 60--62 of Birrell
and Davies~\cite{Birrell-Davies}.  Birrell and Davies were interested
in the problem of particle production engendered by the expansion of
the universe in a cosmological context. Although the physical model is
radically different here the mathematical aspects of the analysis
carry over with some minor translation in the details.  Equations
(3.88) of Birrell--Davies become
\begin{equation}
\omega^\tau_\in  
= k \sqrt{a-b}
= k \sqrt{\epsilon_\in } 
= k \; \ni ;
\end{equation}
\begin{equation}
\omega^\tau_\out  
= k \sqrt{a+b} 
= k \sqrt{\epsilon_\out } 
= k \; \no ;
\end{equation}
\begin{equation}
\omega^\tau_{\pm} = 
\half \; k \; \left| \ni  \pm  \no  \right| = 
\half | \omega^\tau_\in \pm \omega^\tau_\out|.
\end{equation}
(Here we emphasise that these frequencies are those appropriate to the
``pseudo-time'' $\tau$.)  The Bogolubov $\alpha$ and $\beta$
coefficients can be easily deduced from Birrell--Davies (3.92)+(3.93)
\begin{eqnarray}
\alpha(\vec k_\in , \vec k_\out )  
&=& \frac{\sqrt{\omega^\tau_\out \; \omega^\tau_\in }}{\omega^\tau_{+}} \;\;
{\Gamma(-i \omega^\tau_\in  \tau_0) \;
 \Gamma(-i \omega^\tau_\out  \tau_0) \over
 \Gamma(-i \omega^\tau_{-} \tau_0)^2 } \;
\delta^3(\vec k_\in  - \vec k_\out  )\\
\beta(\vec k_\in , \vec k_\out )  
&=& -
\frac{\sqrt{\omega^\tau_\out \; \omega^\tau_\in }}{\omega^\tau_{-}} \;\;
{\Gamma(-i \omega^\tau_\in  \tau_0) \;
 \Gamma(i \omega^\tau_\out  \tau_0) \over
 \Gamma(i \omega^\tau_{-} \tau_0)^2 } \;
\delta^3(\vec k_\in  + \vec k_\out  ).
\end{eqnarray}
Now square, using Birrell--Davies (3.95). We obtain\footnote{
Note that these are the Bogolubov coefficients for a scalar field
theory. For QED in the infinite volume limit the two photon
polarizations decouple into two independent scalar fields and these
Bogolubov coefficients can be applied to each polarization state
independently. Finite volume effects are a little trickier.}
%
\begin{eqnarray}
|\beta(\vec k_\in , \vec k_\out )|^2 
&=& 
{\sinh^2(\pi\omega^\tau_{-}\tau_0)\over
  \sinh(\pi\omega^\tau_\in \tau_0) \;
 \sinh(\pi\omega^\tau_\out \tau_0)} \;
{V\over (2\pi)^3 } \; 
\delta^3(\vec k_\in  + \vec k_\out  ).
\end{eqnarray}
We now need to translate this into physical time, noting that
asymptotically, in either the infinite past or the infinite future,
$t \approx \epsilon \tau + (constant)$, so that for physical
frequencies
\begin{equation}
\omega_\in  = 
{\omega^\tau_\in \over\epsilon_\in } = 
{\omega^\tau_\in \over n_\in ^2} = 
{k \sqrt{a-b}\over \epsilon_\in }  = 
k \sqrt{1\over\epsilon_\in } 
=  {k \over n_\in };
\end{equation}
\begin{equation}
\omega_\out  = 
{\omega^\tau_\out \over\epsilon_\out } = 
{\omega^\tau_\out \over n_\out ^2} = 
{k \sqrt{a+b}\over \epsilon_\out } = 
k \sqrt{1\over\epsilon_\out } 
= {k \over n_\out }.
\end{equation}
Note that there is a symmetry in the Bogolubov coefficients under
interchange of ``in'' and ``out''.  

We also need to convert the timescale over which the refractive index
changes form pseudo-time to physical time. To do this we define
\begin{equation}
t_0 \equiv \tau_0 \left.{dt\over d\tau}\right|_{\tau=0}.
\end{equation}
For the particular temporal profile we have chosen for analytic
tractability this evaluates to
\begin{equation}
t_0 = \half  \tau_0 \left( \ni^2 + \no^2 \right).
\end{equation}
After these substitutions, the (squared) Bogolubov coefficient becomes
\begin{eqnarray}
|\beta(\vec k_\in , \vec k_\out )|^2 
&=& 
{
\sinh^2\left(
\pi\; {\textstyle |\ni^2 \omega_\in -\no^2 \omega_\out| \over\textstyle \ni^2+\no^2} \; t_0
\right)
\over
\sinh\left(
2\pi \; {\textstyle \ni^2 \over \textstyle \ni^2+\no^2} \; \omega_\in t_0
\right) \;
\sinh\left(
2\pi \; {\textstyle \no^2 \over \textstyle \ni^2+\no^2} \; \omega_\out t_0
\right)
} \;
{V\over (2\pi)^3 } \; 
\delta^3(\vec k_\in  + \vec k_\out  ).
\label{bog2}
\end{eqnarray}
We now consider two limits, the adiabatic limit and the sudden limit,
and investigate the physics.

\subsection{Sudden limit}

Take
\begin{equation}
\max \{\omega^\tau_\in , \omega^\tau_\out , \omega^\tau_-\} \; \tau_0 \ll 1.
\end{equation}
This corresponds to a rapidly changing refractive index. In terms of
physical time this is equivalent to
\begin{equation}
2\pi\;\max 
\left\{1, {\ni\over\no}, \half\left|{\ni\over\no}-1\right|\right\} \; 
{ \no^2 \over \ni^2+\no^2} \; \omega_\out t_0  \ll 1,
\end{equation}
which can be simplified to yield
\begin{equation}
2\pi\;\max \{\ni,\no\} \; 
{ \no \over \ni^2+\no^2} \; \omega_\out t_0  \ll 1.
\end{equation}
So the sudden approximation is a good approximation for frequencies
{\em less} than $\Omega_{\mathrm sudden}$, where we define
\begin{equation}
\Omega_{\mathrm sudden} = {1\over 2\pi t_0} \; 
{\ni^2+\no^2\over \no \; \max\{\ni,\no\}
}.
\end{equation}
The this shows that the frequency up to which the sudden approximation
holds is not just the reciprocal of the timescale of the change in the
refractive index: there is also a strong dependence on the initial and
final values of the refractive indices. This implies that we can
relax, for some ranges of values of $\ni$ and $\no$, our figure of
$t_0\sim O({\rm fs})$ by up to a few orders of magnitude.
Unfortunately the precise shape of the spectrum is heavily dependent
on all the experimental parameters ($K,\ni,\no,R$). This
discourages us from making any sharp statement regarding the exact
value of the timescale required in order to fit the data.

In the region where the sudden approximation holds the various
$\sinh(x)$ functions in equation (\ref{bog2}) can be replaced by their
arguments $x$. Then
\begin{equation}
|\beta|^2 \propto 
{ (\pi [\ni-\no])^2 \over 
(2\pi \ni) \; (2\pi \no) }. 
\label{E:sb}
\end{equation}
More precisely
\begin{equation}
|\beta(\vec k_\in, \vec k_\out)|^2 \approx 
{1\over 4} 
{ (\ni - \no)^2 \over 
\ni \; \no }\;
{V\over (2\pi)^3 } \; \delta^3(\vec k_\in + \vec k_\out ),
\label{E:sbsq}
\end{equation}
For completeness we also give the unsquared Bogolubov
coefficients evaluated in the sudden approximation:
\begin{eqnarray}
\alpha(\vec k_\in, \vec k_\out) 
&\approx&
{1\over 2}
{\ni +\no \over \sqrt{\ni \; \no }}\;
\delta^3(\vec k_\in - \vec k_\out ),\\
\beta(\vec k_\in, \vec k_\out) 
&\approx&
{1\over 2}
{|\ni -\no| \over \sqrt{\ni \; \no }}\;
\delta^3(\vec k_\in + \vec k_\out ).
\end{eqnarray}
As expected, for $\ni \rightarrow \no$, we have $\alpha \rightarrow
\delta^3(\vec k_\in - \vec k_\out )$ and $\beta \rightarrow 0$.

These result should be be compared with that obtained in the companion
paper~\cite{Companion}, where we first include finite volume
effects and then consider the large-volume limit for dielectric
bubbles in order to reproduce the original Schwinger estimate of
photon production~\cite{Sc1,Sc2,Sc3,Sc4,Sc5,Sc6,Sc7}. It should
also be compared with the discussion of Yablonovitch~\cite{Yablonovitch}
[see particularly the formulae in the paragraph between equations
(8) and (9)].

\subsection{Adiabatic limit}

Now take
\begin{equation} 
\min\{\omega^\tau_\in , \omega^\tau_\out , \omega^\tau_-\} \; \tau_0 \gg 1.
\end{equation}
This corresponds to a slowly changing refractive index.  In this limit
the $\sinh(x)$ functions in the exact Bogolubov coefficient can be
replaced with exponential functions $\exp(x)$.  Then
\begin{eqnarray}
|\beta|^2 &\propto& 
{ \exp(2\pi \omega^\tau_-  \tau_0) \over 
\exp(\pi\omega^\tau_\in \tau_0) \;
\exp(\pi\omega^\tau_\out \tau_0)}
\\
&=&
{ \exp(\pi \; |\omega^\tau_\in - \omega^\tau_\out|  \; \tau_0) \over 
\exp(\pi\omega^\tau_\in \tau_0) \; 
\exp(\pi\omega^\tau_\out \tau_0)}.
\end{eqnarray}
More precisely 
\begin{equation}
|\beta(\vec k_\in , \vec k_\out )|^2 
\approx 
\exp\left(
-2\pi \; \min\{\omega^\tau_\out ,\omega^\tau_\in \} \; 
\tau_0
\right) \;
{V\over (2\pi)^3 } \; 
\delta^3(\vec k_\in  + \vec k_\out  ).
\end{equation}
In terms of physical time the condition defining the adiabatic limit
reads
\begin{equation} 
2\pi\;\min 
\left\{1, {\ni\over\no}, \half\left|{\ni\over\no}-1\right|\right\} 
\; 
{ \no^2 \over \ni^2+\no^2} \; \omega_\out t_0  \gg 1.
\end{equation}
The Bogolubov coefficient then becomes
\begin{equation}
|\beta(\vec k_\in , \vec k_\out )|^2 
\approx 
\exp\left(-4\pi  \; {\min\{\ni,\no\} \;
\no \over \ni^2 + \no^2} \; \omega_\out t_0
\right) \;
{V\over (2\pi)^3 } \; 
\delta^3(\vec k_\in  + \vec k_\out  ),
\end{equation}
This implies exponential suppression of photon production for
frequencies {\em large} compared to 
\begin{equation}
\Omega_{\mathrm adiabatic} \equiv 
{1\over 2\pi t_0} 
{\ni^2 + \no^2 \over \no \;\min\{\ni,\no,\half|\ni-\no|\} }.
\end{equation}
Eberlein's model~\cite{Eberlein1,Eberlein2,Eberlein3} for
sonoluminescence explicitly makes the adiabatic approximation and this
effect is the underlying reason why photon production is so small in
that model; of course the technical calculations of Eberlein's model
also include the finite volume effects due to finite bubble radius
which somewhat obscures the underlying physics of the adiabatic
approximation.

\subsection{The transition region}

Generally there will be a transition region between $\Omega_{\mathrm
sudden}$ and $\Omega_{\mathrm adiabatic}$ over which the Bogolubov
coefficient has a different structure from either of the asymptotic
limits.  In this transition region the Bogolubov coefficient is well
approximated by a monomial in $\omega$ multiplied by an exponential
suppression factor, but the e-folding rate in the exponential is
different from that in the adiabatic regime. Fortunately, we will not
need any detailed information about this region, beyond the fact that
there is an exponential suppression.

\subsection{Spectrum}

The number spectrum of the emitted photons is
\begin{equation}
\label{E:spectrum0}
{dN(\vec k_\out )\over d^3\vec k_\out }
=
\int|\beta(\vec k_\in ,\vec k_\out )|^{2} 
d^3\vec k_\in .
\end{equation}
Taking into account that $d^3\vec k_\out = 4\pi k_\out^2 
\; d k_\out $ this easily yields 
 \begin{equation}
\label{E:spectrum1} 
{dN(\omega_\out )\over d\omega_\out }
=
{
\sinh^2\left(
{ \textstyle\pi\; |n_\in -n_\out | \; n_\out  \; \omega_\out  t_0
\over
\textstyle (\ni^2+\no^2)
}
\right)
\over
\sinh\left(
{\textstyle2\pi n_\in \; n_\out  \omega_\out  t_0
\over 
\textstyle(\ni^2+\no^2)
}
\right) \;
\sinh\left(
{\textstyle 2\pi n_\out ^{2}\;  \omega_\out  t_0
\over
\textstyle (\ni^2+\no^2)
}
\right)
} \;
{2V\over (2\pi)^3 } \; 
4\pi \omega_\out ^2 \; n_\out ^3 . 
\end{equation}
(Here the factor 2 is introduced by hand by taking into account
the 2 photon polarizations).  For low frequencies (where the sudden
approximation is valid) this is a phase-space limited spectrum with
a prefactor that depends only on the overall change of refractive
index. For high frequencies (where the adiabatic approximation
holds sway) the spectrum is cutoff in an exponential manner depending
on the rapidity of the change in refractive index. 

\begin{figure}[htb]
\vbox{\hfil\psfig{figure=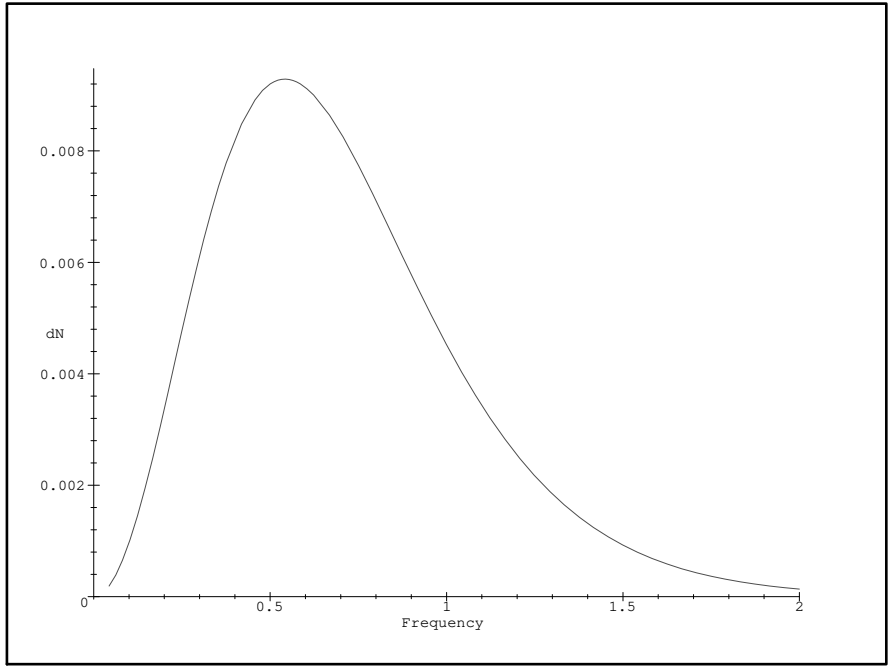,height=10cm,angle=270}\hfil}
\caption{%
Number spectrum (photons per unit volume) for $ n_\in =1$, $ n_\out =2$.  
The horizontal axis is $\omega_\out$ and is expressed in PHz. 
The typical timescale $t_{0}$ is set equal to one $fs$. 
The vertical axis is in arbitrary units.}
\label{F:toy-spectrum-1} 
\end{figure}

\begin{figure}[htb]
\vbox{\hfil\psfig{figure=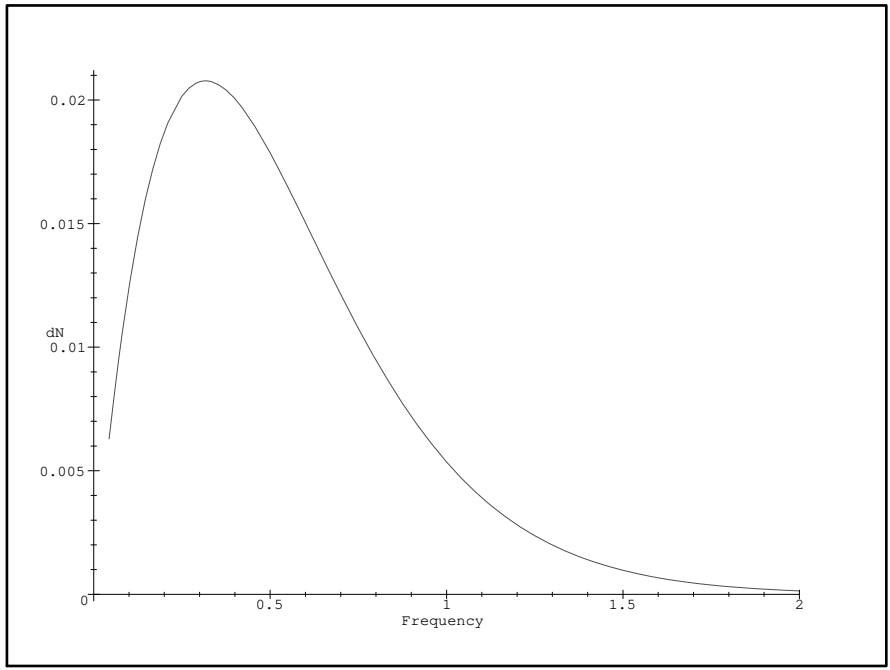,height=10cm,angle=270}\hfil}
\caption{%
Number spectrum for a Planck blackbody curve with $k_B T=
(\ni^2+\no^2)/ (4\pi t_0\; \no \; \min\{\ni,\no,\half|\ni-\no|\} ). $
The horizontal axis is $\omega_\out$ and is expressed in PHz. The
typical timescale $t_{0}$ is set equal to one fs.  The vertical axis
is in arbitrary units (but with the same normalization as figure 1).}
\label{F:toy-spectrum-2} 
\end{figure}

\begin{figure}
[htb]
\vbox{\hfil\psfig{figure=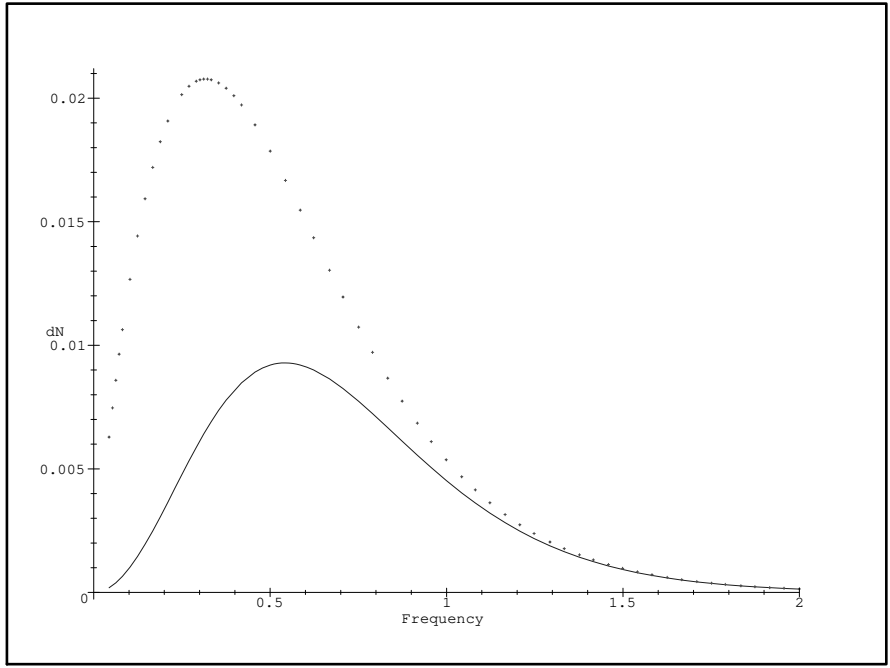,height=10cm,angle=270}\hfil}
\caption{%
Superimposed number spectra (the Planck spectrum is the dotted one).
This figure demonstrates the similar high-frequency behaviour although
low energy behaviour is different (quadratic versus linear).}
\label{F:toy-spectrum-3} 
\end{figure}

A sample spectrum is plotted in figure \ref{F:toy-spectrum-1}.  For
comparison figure \ref{F:toy-spectrum-2} shows a Planckian spectrum
with the same exponential falloff at high frequencies, while the two
curves are superimposed in figure \ref{F:toy-spectrum-3}.


\subsection{Lessons from this toy model}

{\bf Lesson 1:} This is only a toy model, but we feel it adequately
proves that efficient photon production occurs only in the sudden
approximation, and that photon production is suppressed in the
adiabatic regime. The particular choice of profile $\epsilon(\tau)$
was merely a convenience, it allowed us to get analytic exact results,
but it is not a critical part of the analysis. One might worry that
the results of this toy model are specific to the choice of profile
(\ref{E:profile}). That the results are more general can be
established by analyzing general bounds on the Bogolubov coefficients,
which is equivalent to studying general bounds on one-dimensional
potential scattering~\cite{Visser}. We shall here quote only the key
result that for any monotonic change in the dielectric constant the
sudden approximation provides a strict upper bound on the magnitude of
the Bogolubov coefficients~\cite{Visser}.

{\bf Lesson 2:} Eberlein's model for
sonoluminescence~\cite{Eberlein1,Eberlein2,Eberlein3} explicitly
uses the adiabatic approximation. For arbitrary adiabatic changes
we expect the exponential suppression to still hold with $\rho$
now being some measure of the timescale over which the refractive
index changes.

{\bf Lesson 3:} Schwinger's model for
sonoluminescence~\cite{Sc1,Sc2,Sc3,Sc4,Sc5,Sc6,Sc7} implicitly uses
the sudden approximation.  It is only for the sudden approximation
that we recover Schwinger's phase-space limited spectrum.  For
arbitrary changes the sudden approximation provides a rigorous
upper bound on photon production. It is only in the sudden
approximation that efficient conversion of zero-point fluctuations
to real photons takes place.  Though this result is derived here
only for a particularly simple toy model we expect this part of
the analysis to be completely generic. We expect that any mechanism
for converting zero-point fluctuations to real photons will exhibit
similar effects.

\subsection{Extensions of this toy model}

The major weaknesses of the toy model are that it currently includes
neither dispersive effects nor finite volume effects. Including
dispersive effects amounts to including condensed matter physics
by letting the refractive index itself be a function of frequency.
To do this carefully requires a very detailed understanding of the
condensed matter physics, which is quite beyond the scope of the
present paper. Instead, in this section we shall content ourselves with
making order-of-magnitude estimates using Schwinger's sharp cutoff
for the refractive index and the sudden approximation.

The second issue, that of finite volume effects, is addressed more
carefully in the companion paper~\cite{Companion}. Finite volume
effects are expected to be significant but not overwhelmingly large.
{From} estimates of the available Casimir energy developed
in~\cite{MV}, the fractional change in available Casimir energy
due to finite volume effects is expected to be of order $1/(K R)
= \hbox{(cutoff wavelength)} /(2\pi \hbox{(minimum bubble radius)})$
which is approximately $(300\; \hbox{nm})/(2\pi \;500\; \hbox{nm})
\approx 10\%\;$\footnote{
Here we have estimated the cutoff wavelength from the location of
the peak in the SL spectrum. If anything, this causes us to overestimate
the finite volume effects.}.

Returning to dispersive issues: if the refractive indices 
were completely non-dispersive (frequency-independent), then the sudden
approximation would imply infinite energy production. In the real
physical situation $\ni$ is a function of $\omega_\in $
and $\no$ is a function of $\omega_\out $.  Schwinger's
sharp momentum-space cutoff for the refractive index is equivalent,
in this formalism, to the choice
\begin{equation}
\ni(k) = \ni \; \Theta(K_\in -k) + 1 \; \Theta(k-K_\in),
\end{equation}
\begin{equation}
\no(k) = \no \; \Theta(K_\out -k) + 1 \; \Theta(k-K_\out),
\end{equation}
(More complicated models for the cutoff are of course possible at the
cost of obscuring the analytic properties of the model) Although in
general the two cutoff wavenumbers, $K_\in$ and $K_\out$ can be
different, it is easy to show (using the delta function and the fact
that $\beta \rightarrow 0$ when $\ni=\no=1$), that this is equivalent
to a single common cutoff $K\equiv \min\{K_\in,K_\out\}$.  {From}
equation (\ref{E:sbsq}), taking into account the two photon
polarizations, one obtains 
\begin{eqnarray}
|\beta(\vec k_\in ,\vec k_\out )|^{2}
&\approx&
{1 \over 2} \frac{\left(\no-\ni\right)^2}{\ni \no}
{V\over(2\pi)^3}\; 
\Theta(K - k_\in ) \;  \Theta(K - k_\out ) \;
\delta^3(\vec k_\in  + \vec k_\out ).
\label{E:largeb2}
\end{eqnarray}
As a consistency check, expression (\ref{E:largeb2}) has the desirable
property that $\beta\to0$ as $\no\to\ni$: That is, if there is no
change in the refractive index, there is no particle production.  In
fact the computed Bogolubov coefficient is directly related to the
physical quantities we are interested in
\begin{equation}
{dN\over d^3 \vec k_\out}
=\int|\beta(\omega_\in ,\omega_\out )|^{2}
\; d^3 \vec k_\in,
\end{equation}
\begin{equation}
{dN\over d k_\out}
=4\pi k_\out^2 \; \int|\beta(\omega_\in ,\omega_\out )|^{2}
\; d^3 \vec k_\in,
\end{equation}
\begin{equation}
{dN\over d\omega_\out}
=4\pi \frac{\no^3 \omega_\out^2}{c^3} \; 
\int|\beta(\omega_\in ,\omega_\out )|^{2}
\; d^3 \vec k_\in,
\end{equation}
\begin{equation} 
N=\int {dN\over d\omega_\out} \; d\omega_\out ,
\end{equation}
and
\begin{equation}
E= \hbar \int {dN(\omega_\out )\over d\omega_\out }   
\; \omega_\out  \; d\omega_\out .
\end{equation}
So we can now compute the spectrum, the number, and the total energy
of the emitted photons.
\begin{eqnarray}
{dN(\omega_\out )\over d\omega_\out }
=
\frac{\no}{c} \; {dN(\omega_\out )\over dk_\out }
&=&
\frac{\no}{c} \; 4\pi k^{2}_\out {dN(\omega_\out )\over d^3{\vec k}_\out }
\\
&\approx&
{\no \over (2\; c)}\;
{\left(\no-\ni\right)^2\over\no\,\ni}
{V\over(2\pi)^3}\; 4\pi \; k_\out ^2 \;
\Theta(K - k_\out ) \;
\\
&=&
{1\over (2\; c^3)}\;{\no^2} \;
{\left(\no-\ni\right)^2\over\ni}
{V\over(2\pi)^3}\; 4\pi \; \omega_\out ^2 \;
\Theta\left(K - \frac{\no \omega_\out}{c}\right)
\end{eqnarray}
The number of emitted photons is then approximately
\begin{eqnarray}
N
&\approx&
{1\over 2} \;{\no^2} \;
{\left(\no-\ni\right)^2\over\ni} \;
{V\over(2\pi)^3}\; {4\pi\over3} \; (K/\no)^3
\\
&=&
{1\over 12\pi^2} \;
{\left(\no-\ni\right)^2\over\ni\no} \;
{V K^3}.
\end{eqnarray}
So that for a spherical bubble  
\begin{eqnarray}
N
&\approx&
{1\over 9\pi} \;
{\left(\no-\ni\right)^2\over\no\ni} \;
(R K)^3.
\end{eqnarray}
It is important to note that the wavenumber cutoff $K$ appearing in
the above formula is not equal to the observed wavenumber cutoff
$K_\observed$. The observed wavenumber cutoff is in fact the upper
wavelength measured once the photons have left the bubble and entered
the ambient medium (water), so actually
\begin{equation}
K={\omega_{\mathrm max}\no\over c}={\no\over\nl} \; K_\observed.
\end{equation}
Thus
\begin{eqnarray}
N
&\approx&
{1\over 9\pi} \;
{\left(\no-\ni\right)^2\over\no\ni} \;
\left(R \;{\no\over\nl} \; K_\observed\right)^3.
\\
&=&
{1\over 9\pi} \;
{\left(\no-\ni\right)^2\over\ni} \; \no^2 \; 
\left({R \omega_{\mathrm max} \over c}\right)^3.
\end{eqnarray}
The total emitted energy is approximately
\begin{eqnarray}
E&\approx&
{1\over 2} \; \frac{\no^2}{c^3} \;
{\left(\no-\ni\right)^2\over\ni}
{V\over(2\pi)^3}\; 4\pi \;
\int \hbar \omega_\out  \; \omega_\out ^2 \;
\Theta\left(K - \frac{\no \omega_\out}{c} \right) d\omega_\out 
\\
&=&
{\hbar\over 2} \; \frac{\no^2}{c^3} \;
{\left(\no-\ni\right)^2\over\ni} \;
{V\over(2\pi)^3}\; {4\pi\over4} \; (K\;c/\no)^4
\\ 
&=&
{1\over16\pi^2} \;
{\left(\no-\ni\right)^2\over\ni\no^2} \;
\hbar\; c\; K \; {V K^3}
\\
&=&
\frac{3}{4}\; N\; \hbar \omega_{\mathrm max}.
\label{E:energy}
\end{eqnarray}
So the average energy per emitted photon is approximately\footnote{%
The maximum photon energy is $\hbar \; \omega_{\mathrm max} \approx 4\;
{\rm eV}$.  }
%
\begin{equation}
\langle E \rangle = 
{3\over 4} \hbar c\; K/\no = {3\over 4} \hbar  \; \omega_{\mathrm
max}\sim 3 \; {\rm eV}.
\end{equation}

Taking into account this extra factor we can now consider some  numerical 
estimates based on our results.

\subsection{Some numerical estimates}

In Schwinger's original model he took $\ngi\approx 1$, $\ngo \approx
\nl \approx 1.3$, $V= (4\pi/3) R^3$, with $R \approx R_{\mathrm
max} \approx 40 \; \mu {\rm m}$ and  $K\approx 2\pi/(360\; {\rm
nm})$ \cite{Sc4}.  Then $KR \approx 698$. Substitution of these
numbers into equation (\ref{E:schwinger}) leads to an energy budget
suitable for about {\em three} million emitted photons.

By direct substitution in equation (\ref{E:energy}) it is easy to
check that Schwinger's results can qualitatively be recovered also in
our formalism: in our case we get about {\em 1.8 million} photons for
the same numbers of Schwinger and about {\em 4 million} photons using
the updated experimental figures $R_{\mathrm max} \approx 45 \;
\mu {\rm m}$ and  $K\approx 2\pi/(300\; {\rm nm})$. 

A sudden change in refractive index {\em would} indeed convert the most 
of the energy budget based on static Casimir energy calculations into 
real photons. This may be interpreted as an independent check on Schwinger's
estimate of the Casimir energy of a dielectric sphere.  Unfortunately,
the sudden (femtosecond) change in refractive index required to
get efficient photon production is also the fly in the ointment
that kills Schwinger's original choice of parameters:  The collapse
from $R_{\mathrm max}$  to $R_{\mathrm min}$ is known to require
approximately $10\; {\rm ns}$, which is far too long a timescale
to allow us to adopt the sudden approximation.

In our new version of the model we have $R\approx R_{\mathrm
light-emitting-region} \approx R_{\mathrm min} \approx 500 \;
\hbox{nm} $ and take $K_\observed\approx 2\pi/(200\; {\mathrm nm})$ so
that $K_\observed R\approx 5 \pi \approx 15$. To get about one million
photons we now need, for instance, $\ni\approx 1$ and $\no\approx 12$,
or $\ni\approx 2\times 10^4$ and $\no\approx 1$, or even $\no \approx
25$ and $\ni\approx 71$, though many other possibilities could be
envisaged.  In particular, the first set of values could correspond to
a change of the refractive index at the van der Waals hard core due to
a sudden compression {\em e.g.}, generated by a shock wave. In this
framework it is obvious that the most favorable composition for the
gas would be a noble gas since this mechanism would be most effective
if the gas could be enormously compressed without being easily
ionizable\footnote{%
To ionize Argon requires $15 \; {\rm eV}\approx 10^5 \; {\rm K}/k_{B}$
per atom. This energy could be provided either from a heat bath at
this temperature (Bremsstrahlung) or from kinetic energy given by
atomic collisions. Both of these possibilities require very extreme
hypotheses.}.

Note that the estimated values of $\ngo$ and $\ngi$ are extremely
sensitive to the precise choice of cutoff, and the size of the light
emitting region, and that the approximations used in taking the
infinite volume limit underlying the use of our homogeneous dielectric
model are uncontrolled. (The complications attendant on any attempt at
including finite volume effects are sufficiently complex as to warrant
being relegated to a separate technical paper.~\cite{Companion}) We
should not put to much credence in the particular numerical value of
$\ngo$ estimated by these means, but should content ourselves with
this qualitative message: We need the refractive index of the contents
of the gas bubble to change dramatically and rapidly to generate the
photons. Compare this with the calculation and arguments presented by
Yablonovitch~\cite{Yablonovitch}, who points out that ionization
processes can and often do cause such sudden drops in the refractive
index.

As a final remark we stress that equation (\ref{E:schwinger}) and
equation (\ref{E:energy}) are not quite identical. The volume term for
photon production that we have just derived [equation
(\ref{E:energy})] is of second order in $(\ni-\no)$ and not of first
order like equation (\ref{E:schwinger}). This is ultimately due to the
fact that the interaction term responsible for converting the initial
energy in photons is a pairwise squeezing operator
(see~\cite{2gamma}).  Equation (\ref{E:energy}) demonstrates that any
argument that attempts to deny the relevance of volume terms to
sonoluminescence due to their dependence on $(\ni-\no)$ has to be
carefully reassessed.  In fact what you measure when the refractive
index in a given volume of space changes is {\em not} directly the
change in the static Casimir energy of the ``in'' state, but rather
the fraction of this static Casimir energy that is converted into
photons. We have just seen that once conversion efficiencies are taken
into account, the volume dependence is conserved, but not the power in
the difference of the refractive index.  Indeed the dependence of
$|\beta|^2$ on $(\ni-\no)^2$, and the symmetry of the former under the
interchange of ``in'' and ``out'' states, also proves that it is the
amount of change in the refractive index and not its ``direction''
that governs particle production.  This apparent paradox is easily
solved by taking into account that the main source of energy is the
acoustic field and that the amount of this energy actually converted
in photons during each cycle is a very small fraction of the total
acoustic energy.

\subsection{Estimating the number of photons}

Using the above as a guide to the appropriate starting point, we can
now systematically explore the relationship between the in and out
refractive indexes and the number of photons produced. Using
$K_\observed R \approx 15$ we get
\begin{equation}
N = {119\over\nl^3}\; (\no-\ni)^2 \; {\no^2 \over \ni}.
\end{equation}

\begin{figure}[htb]
\vbox{\hfil\psfig{figure=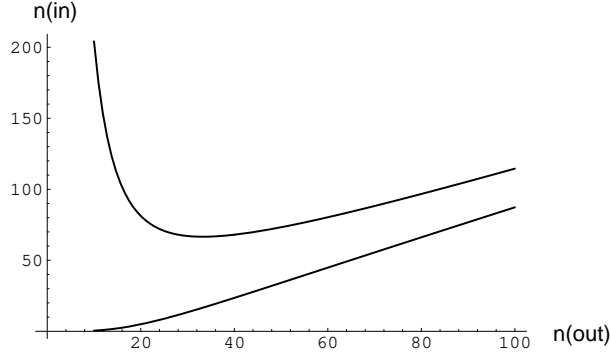,height=5cm}\hfil}
\caption{%
The initial refractive index $\ni$ plotted as a function of $\no$ when
one million photons are emitted in the sudden approximation.}
\label{F:photons-1} 
\end{figure}

This equation can be algebraically solved for $\ni$ as a function of
$\no$ and $N$. (It's a quadratic.) For $N=10^6$ emitted photons the
result is plotted in figure (\ref{F:photons-1}). For any specified
value of $\no$ there are exactly two values of $\ni$ that lead to one
million emitted photons.  To understand the qualitative features of
this diagram we consider three sub-regions.

\begin{figure}[htb]
\vbox{\hfil\psfig{figure=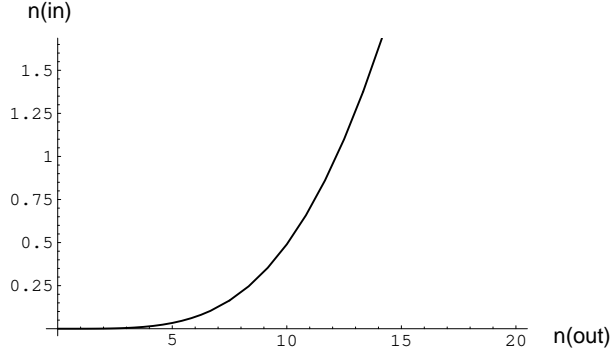,height=5cm}\hfil}
\caption{%
The initial refractive index $\ni$ plotted as a function of $\no$ when
one million photons are emitted in the sudden approximation. Here we
focus on the branch that approaches the origin.}
\label{F:photons-2} 
\end{figure}

First, if $\ni \ll \no$ then we can approximate
\begin{equation}
\ni \approx {119 \; \no^4 \over \nl^3 \; N}.
\end{equation}
This corresponds to the region near the origin, and we focus on this
region in figure (\ref{F:photons-2}).

\begin{figure}[htb]
\vbox{\hfil\psfig{figure=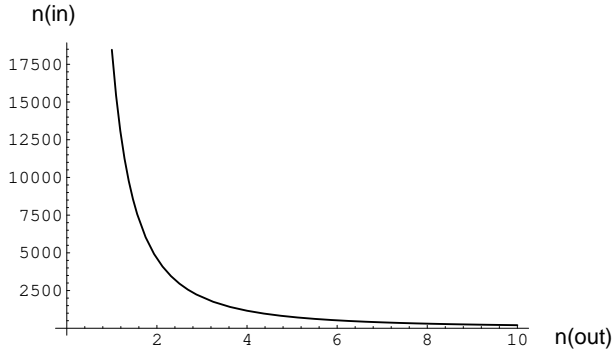,height=5cm}\hfil}
\caption{%
The initial refractive index $\ni$ plotted as a function of $\no$ when
one million photons are emitted in the sudden approximation. Here we
focus on the branch that approaches the $y$ axis.}
\label{F:photons-3} 
\end{figure}

Second, if $\ni \gg \no$ then we can approximate
\begin{equation}
\ni \approx {\nl^3 \; N \over 119 \; \no^2}.
\end{equation}
This corresponds to the region near the $y$ axis, and we focus on this
region in figure (\ref{F:photons-3}).

Third, if $\ni \approx \no$ then we can approximate
\begin{equation}
N \approx {119\over \nl^3} \; (\ni - \no)^2 \; \no,
\label{nn}
\end{equation}
so that
\begin{equation}
\ni \approx \no \pm \sqrt{N \; \nl^3 \over 119 \; \no}.
\label{nnn}
\end{equation}
This corresponds to the region near the asymptote $\ni=\no$.

Thus to get a million photons emitted from the van der Waals hard core
in the sudden approximation requires a significant (but not enormous)
change in refractive index. There are many possibilities consistent
with the present model and the experimental data.

\section{Experimental features and possible tests}

Our proposal shares with other proposals based on the dynamical
Casimir effect the main points of strength previously sketched.  On
the other hand we feel it important to stress that the model we have
developed implies a much more complex and rich collection of physical
effects due to the fact that photon production from vacuum is no
longer due to the simple motion of the bubble boundary. The model
indicates that a viable Casimir route to SL cannot avoid a ``fierce
marriage'' with features related to condensed matter physics.  As a
consequence our proposal is endowed both with general characteristics,
coming from its Casimir nature, and with particular ones coming from
the details underlying the sudden change in the refractive index.

Although the calculation presented above is just a ``probe'', we
can see that it is already able to make some general predictions
that one can expect to see confirmed in a more complete approach.
First of all the photon number spectrum the model predicts is not
a black body. It is polynomial at low frequencies ($\omega^{2}$ in
the infinite volume approximation of this paper), and in principle
this difference can be experimentally detected.  (The same qualitative
prediction can be found in \Schutzhold\ {\em et al.} \cite{SPS}.)
Moreover the spectrum is expected to be a power law dramatically
ending at frequencies corresponding to the physical wave-number
cutoff $K$ (at which the refractive indices go to 1). This cutoff
implies the absence of hard UV photons and hence, in accordance
with experiments, the absence of dissociation phenomena in the
water surrounding the bubble.

It is the sudden approximation adopted in this paper that makes it
possible to mimic the experimentally observed spectrum.  For either a
rise in refractive index from $1$ to $12$, or a drop in refractive
index from $2\times 10^4$ to $1$ one can produce approximately one
million photons with frequency mainly in the visible range.  The
quasi-thermal nature of the emitted photons can be explained by the
squeezed nature of the photon pairs that are generically created via
dynamical Casimir effect (see reference \cite{2gamma}).  Single-photon
measurements are then thermal but the core of the bubble is not
required to achieve the tremendous physical temperatures envisaged by
other models. The apparent temperature measured in single-photon
observables can be instead linked to the degree of squeezing of the
photon-pairs.  As it will be explained in \cite{2gamma}, two-photon
observables do not exhibit the same thermal statistics, and therefore
measurements of suitable two-photon observables provide a useful
diagnostic for SL models based on the dynamical Casimir effect.  This
is a general feature of all models based on photon creation from the
QED vacuum and hence it can be used as a definitive test of the
presence of a dynamical Casimir effect.

In this type of model, the flash of photons is predicted to occur at
the end of the collapse, the scale of emission zone is of the order of
$500\; \hbox{nm}$, and the timescale of emission is very short (with a
rise-time of the order of femtoseconds, though the flash duration may
conceivably be somewhat longer~\footnote{It would be far too naive to
assume that femtosecond changes in the refractive index lead to pulse
widths limited to the femtosecond range. There are many condensed
matter processes that can broaden the pulse width however rapidly it
is generated. Indeed, the very experiments that seek to measure the
pulse width \cite{Flash1,Flash2} also prove that when calibrated with
laser pulses that are known to be of femtosecond timescale, the SL
system responds with light pulses on the picosecond timescale.}.
These are points in substantial agreement with observations. In the
infinite volume limit the photons emerge in strictly back-to-back
fashion. In contrast, for a finite volume bubble we have shown
in~\cite{Companion} that the size of the emitting region constrains
the model to low angular momentum for the out states. This is a very
sharp prediction that is in principle testable with a suitable
experiment devoted to the study of the angular momentum decomposition
of the outgoing radiation.

Regarding other experimental dependencies, such as the temperature of
the water or the role of noble gases, we can give general arguments
but a truly predictive analysis can be done only after focussing on a
specific mechanism for changing the refractive index.

For instance, the presence of noble gas is likely to change
solubilities of gas in the bubble, and this can vary both bubble
dynamics and the sharpness of the boundary.  Alternatively, a small
percentage of noble gas in air can be very important in the behavior
of its dielectric constant at high pressure. Indeed, while small
admixtures of noble gas will not significantly alter the
zero-frequency refractive index, from the Casimir point of view the
behaviour of the refractive index over the entire frequency range up
to the cutoff is important.

Finally, the temperature of water can instead affect the dynamics of
the bubble boundary by influencing the stability of the bubble,
changing either the solubility of air in water or the surface tension
of the latter. As observed by
Schwinger~\cite{Sc1,Sc2,Sc3,Sc4,Sc5,Sc6,Sc7}, temperature can also
affect the dielectric cutoff, and so temperature dependence of SL is
quite natural in Schwinger-like approaches.

\section{Discussion and Conclusions}

The present paper presents calculations of the Bogolubov coefficients
relating the two QED vacuum states appropriate to changes in the
refractive index of a dielectric bubble.  We have verified by explicit
computation that photons are produced by rapid changes in the
refractive index, and are in agreement with Schwinger in that QED vacuum
effects remain a viable candidate for explaining SL. However, some
details of the particular model considered in the present paper are
somewhat different from that originally envisaged by Schwinger. Based
largely on the fact that efficient photon production requires
timescales of the order of femtoseconds we were led to consider rapid
changes in the refractive index as the gas bubble bounces off the van
der Waals hard core. It is important to realize that the speed of sound
in the gas bubble can become relativistic at this stage.

A key lesson learned from this paper is that in order that the
conversion of zero-point fluctuations to real photons be relevant for
sonoluminescence we would want the sudden approximation to hold for
photons all the way out to the cutoff ($200\;\hbox{nm}$; corresponding
to a period of $0.66 \times 10^{-15} \; \hbox{seconds}$).  That's a
{\em femtosecond} timescale. This implies that if conversion of
zero-point fluctuations to real photons is a significant part of the
physics of sonoluminescence then the refractive index must be changing
significantly on femtosecond timescales.  {\em Thus the changes in
refractive index cannot be just due to the motion of the bubble wall.}
(The bubble wall is moving at most at Mach 4~\cite{Physics-Reports},
for a $1\;\mu\hbox{m}$ bubble this gives a collapse timescale of
$10^{-10}$ seconds, about 100 picoseconds.)  In this regard, the
comments of Yablonovitch~\cite{Yablonovitch} are particularly
useful. Yablonovitch points out that, for example, sudden ionization
of a gas can lead to substantial changes in the refractive index on
the sub-picosecond timescale.  Nevertheless we do not necessarily
commit ourselves to ionization as being the relevant process in
sonoluminescence, and are quite content with any rapid change in
refractive index, however generated.

This suggests a slightly different physical model from Schwinger's
original suggestion: Certainly the Casimir energy changes as the
bubble collapses, but it is only in the sudden approximation that we
can justify converting almost all of the change in Casimir energy to
real photons. We thus suggest that one should not be focusing on the
actual collapse of the bubble, but rather the way in which the
refractive index changes as a function of space and time: As the
bubble collapses the gases inside are compressed, and although the
refractive index for air (plus noble gas contaminants) is 1 at STP it
should be no surprise to see the refractive index of the trapped gas
undergoing major changes during the collapse process---especially near
the moment of maximum compression when the molecules in the gas bounce
off the van der Waals' hard core repulsive potential.

Thus attempts at using the dynamical Casimir effect to explain SL are
now much more tightly constrained than previously. We have shown that
any plausible model using the dynamical Casimir effect to explain SL
must use the sudden approximation, and must have very rapid changes in
the refractive index with a timescale of femtoseconds.

If the light is being emitted only at the core bounce, at $R\approx
500 \; {\rm nm}$, then we can get the timescales we want
(femtoseconds) without superluminal effects, but we are rather limited
in the amount of angular momentum we can get out.  Of course, the
present model is nowhere near a complete theory: Presently we can
relate only the asymptotic ``in'' states to the asymptotic ``out''
states via these Bogolubov coefficients. A complete theory of SL will
need to address {\em much more} specific timing information and this
will require a fully dynamical approach (from the QFT point of view)
and a deeper understanding (from the condensed matter side) of the
precise spatio-temporal dependence of the refractive index as the
bubble collapses. In absence of such a more detailed description the
present calculation is a useful first step. Moreover it allows us to
specify certain basic ``signatures'' of the effect that may be
amenable to experimental test. To this end we have developed use of
two-photon statistics as a diagnostic for the dynamical Casimir
effect~\cite{2gamma}. In this paper we have addressed the basic
physical scenario; all technical complications due to finite volume
effects are relegated to a companion paper~\cite{Companion}.  We feel
that, as an explanation for sonoluminescence, we have now driven
models based on the dynamical Casimir effect into a relatively small
region of parameter space, and are hopeful of experimental
verification (or falsification) in the not too distant future.

Stripped to its fundamentals, we therefore view the Schwinger
mechanism as this: Bubble collapse leads to changes in the
spatio-temporal distribution of the refractive index, both via
physical movement of the dielectrics, and through the time-dependent
properties of the dielectrics. Changes in the refractive index drive
changes in the distribution of zero-point modes, and this change in
zero-point modes is reflected in real photon production.

In light of these observations we think that one can also derive a
general conclusion about the long-standing debate on the actual value
of the static Casimir energy and its relevance to sonoluminescence:
{\em Sonoluminescence cannot be directly related to the static Casimir
effect}.  (The static Casimir effect is relevant only insofar as it
gives an approximate value for the energy budget).  We hope that the
investigation of this paper will convince everyone that only models
dealing with the actual mechanism of particle creation (a mechanism
which must have the qualities we have discussed) will be able to
eventually prove or disprove the pertinence of the physics of the
quantum vacuum to Sonoluminescence.  This implies that continuing
debate about the static Casimir effect can be now seen as marginal and
irrelevant with respect to the real physical problems of SL.

In conclusion the present calculation (limited though it may be)
represents an important advance: There now can be no doubt that
changes of the refractive index of the gas inside the bubble lead to
production of real photons---the controversial issues now move to
quantitative ones of precise fitting of the observed experimental
data.  We are hopeful that more detailed models and data fitting will
provide better explanations of the details of the SL effect, and
specifically wish to assert that models based on the QED vacuum remain
viable.

\section*{Acknowledgments}

This research was supported by the Italian Ministry of Science (DWS,
SL, and FB), and by the US Department of Energy (MV). MV particularly
wishes to thank SISSA (Trieste, Italy) and Victoria University (Te
Whare Wananga o te Upoko o te Ika a Maui; Wellington, New Zealand) for
hospitality during various stages of this work.  SL wishes to thank
Washington University for its hospitality. DWS and SL wish to thank
E.~Tosatti for useful discussions. SL wishes to thank M.~Bertola and
B.~Bassett for comments and suggestions.  SL also wishes to thank
R.~\Schutzhold\ and G.~Plunien for illuminating discussion on the
relation between their results and Eberlein's calculations.  All
authors wish to thank G.~Barton for his interest and encouragement.
Finally, the comments and interest of K.~Milton are appreciated.


\end{document}